\documentclass[a4paper,10pt]{article}
\usepackage{jheppub} % for details on the use of the package, please
\usepackage[T1]{fontenc} % if needed
\usepackage{bbold}
\usepackage{amsmath,mathtools}
\usepackage{empheq}
\usepackage{booktabs}
\usepackage{xspace}
\DeclareRobustCommand{\ensuremathrm}[1]{\ensuremath{\mathrm{#1}}\xspace}

\DeclareRobustCommand{\rd}{\ensuremathrm{d}} % differential operator
\DeclareRobustCommand{\re}{\ensuremathrm{e}} % Euler e
\DeclareRobustCommand{\GeV}{\ensuremathrm{GeV}\xspace}
\DeclareRobustCommand{\TeV}{\ensuremathrm{TeV}\xspace}

\DeclareRobustCommand{\nnlojet}{\mbox{\textsc{NNLOjet}}\xspace}
\DeclareRobustCommand{\jet}{\text{jet}\xspace}
\DeclareRobustCommand{\yll}{\ensuremath{Y_{\ell\ell}}\xspace}
\DeclareRobustCommand{\mll}{\ensuremath{M_{\ell\ell}}\xspace}
\DeclareRobustCommand{\phs}{\ensuremath{\phi^{*}_\eta}\xspace}

\newcommand{\order}[1]{\mathcal{O}\left(#1\right)}

\newcommand{\as}{\alpha_\mathrm{s}}

\newcommand{\NNNLL}{\text{N${}^3$LL}}
\newcommand{\eps}{\delta}

\newcommand{\pth}{p_{t}^{\rm H}}
\newcommand{\ptz}{p_{t}^{\rm Z}}
\newcommand{\ptgo}{p_{t}^{\gamma_1}}
\newcommand{\ptgt}{p_{t}^{\gamma_2}}
\newcommand{\ptgg}{p_{t}^{\gamma\gamma}}

\newcommand{\etag}{\eta^{\gamma_i}}

\newcommand{\dZ}{\rd\mathcal{Z}[\{{\tilde R}', k_i\}]}

\usepackage{xcolor}

\usepackage{placeins}

\title{\boldmath Fiducial distributions in Higgs and Drell-Yan production at \NNNLL+NNLO}

\author[a]{Wojciech Bizo\'{n},}
\author[b]{Xuan Chen,}
\author[b,c]{Aude Gehrmann-De Ridder,}
\author[b]{Thomas Gehrmann,}
\author[d]{Nigel Glover,}
\author[e]{Alexander Huss,}
\author[e]{Pier Francesco Monni,}
\author[e,f]{Emanuele Re,}
\author[a]{Luca Rottoli,}
\author[g]{Paolo Torrielli.}

\affiliation[a]{Rudolf
  Peierls Centre for Theoretical Physics, University of Oxford, Clarendon Laboratory, Parks Road, Oxford OX1 3PU}
\affiliation[b]{Department of Physics, University of Z\"urich, CH-8057 Z\"urich, Switzerland}
\affiliation[c]{Institute for Theoretical Physics, ETH, CH-8093 Z\"urich, Switzerland}
\affiliation[d]{Institute for Particle Physics Phenomenology, Department of Physics, University of Durham, Durham, DH1 3LE, UK}
\affiliation[e]{CERN, Theoretical Physics Department, CH-1211 Geneva 23, Switzerland}
\affiliation[f]{LAPTh, CNRS, Universit\'e Savoie Mont Blanc, 74940 Annecy, France}
\affiliation[g]{Dipartimento di Fisica and Arnold-Regge Center, Universit\`a di Torino,
and INFN, Sezione di Torino, Via P. Giuria 1, I-10125 Torino, Italy}

\abstract{The perturbative description of certain differential
  distributions across a wide kinematic range requires the matching of
  fixed-order perturbation theory with resummation of large
  logarithmic corrections to all orders. We present precise matched
  predictions for transverse-momentum distributions in Higgs boson (H) and
  Drell-Yan pair (DY) production as well as for the closely related
  $\phs$ distribution at the LHC.
  The calculation is exclusive in the Born kinematics, and allows for
  arbitrary fiducial selection cuts on the decay products of the
  colour singlets, which is of primary relevance for experimental
  analyses.
  Our predictions feature very small residual scale uncertainties and
  display a good convergence of the perturbative series. A comparison
  of the predictions for DY observables to experimental data at $8~\TeV$
  shows a very good agreement within the quoted errors.}


\preprint{\\CERN-TH-2018-105, IPPP/18/34, LAPTH-015/18, OUTP-17-19P, ZU-TH 17/18}

\allowdisplaybreaks

\begin{document} 
\maketitle
\flushbottom

\section{Introduction}
\label{sec:intro}

The accurate prediction and measurement of differential distributions
is of primary importance for the LHC precision programme,
especially in view of the absence of clear signals of new physics
in the data collected so far.
In this context, a special role is played by the kinematic
distributions of a colour singlet produced in association with QCD
radiation. These observables are often measured by reconstructing the decay
products of the colour singlet (whenever possible), which are
sensitive to the accompanying hadronic activity only through kinematic
recoil. As a consequence, measurements of transverse and angular
observables often lead to small experimental systematic
uncertainties~\cite{Khachatryan:2016nbe,Aaij:2016mgv,Aaboud:2017soa,Aaboud:2017oem,Sirunyan:2017igm,Sirunyan:2017exp,Aaboud:2018xdt,Sirunyan:2018ouh}.

The implication of these precise measurements is twofold. On one hand,
they can be used to fit the parameters of the SM Lagrangian
(e.g.~strong coupling constant, or masses) or to calibrate the models
that typically enter the calculation of hadron-collider observables
(like for instance collinear parton distribution functions
(PDFs)~\cite{Ball:2017nwa}, or non-perturbative corrections and
transverse-momentum-dependent
PDFs~\cite{Scimemi:2016ffw,Scimemi:2017etj,Bacchetta:2017gcc}). An
example is given by the differential distributions in $Z$- and $W$-boson
production, that recently were exploited to perform very precise
extractions of the $W$-boson mass~\cite{Aaboud:2017svj} and to
constrain the behaviour of some PDFs~\cite{Boughezal:2017nla}.
On the other hand, an excellent control over kinematic distributions
is a way to set compelling constraints on new-physics models that
would lead to mild shape
distortions. An example is given by the sensitivity of the Higgs
transverse-momentum ($p_t$) distribution to modification of the Yukawa
couplings of the Higgs to quarks~\cite{Bishara:2016jga,Soreq:2016rae}.

In this article we present state-of-the-art predictions for a
class of differential distributions both in Higgs boson (H) and Drell-Yan pair
(DY) production. Specifically, we combine fixed-order calculations at
next-to-next-to-leading order (NNLO) with the recently-obtained
resummation of Sudakov logarithms to
next-to-next-to-next-to-leading-logarithmic order (N$^3$LL), for the
transverse-momentum spectrum of the colour singlet, as well as for the
angular variable $\phs$~\cite{Banfi:2010cf}. In the following, for
simplicity, we will collectively denote $p_t/M$ or $\phs$ by $v$, with
$M$ representing the invariant mass of the colour singlet.

Inclusive and differential distributions for Higgs-boson production in
gluon fusion are nowadays known with very high precision.  The
inclusive cross section has been computed to
next-to-next-to-next-to-leading-order (N$^3$LO) accuracy in
QCD~\cite{deFlorian:1999zd,Harlander:2002wh,Anastasiou:2002yz,Ravindran:2003um,Ravindran:2002dc,Anastasiou:2016cez,Mistlberger:2018etf}
in the heavy-top-quark limit. The impact of all-order effects due to a
combined resummation of threshold and high-energy logarithms has been
studied in detail, and at the current collider energies the
corrections amount to a few-percent of the total cross
section~\cite{Bonvini:2018ixe}, indicating that the missing
higher-order contributions are under good theoretical control. The
state-of-the-art results for the Higgs transverse-momentum spectrum in
fixed-order perturbation theory are the next-to-next-to-leading-order
(NNLO) computations of
Refs.~\cite{Boughezal:2015dra,Boughezal:2015aha,Caola:2015wna,Chen:2016zka},
which have been obtained in the heavy-top-quark limit.  The effect of
finite quark masses on differential distributions at next-to-leading
order has been recently computed in
Refs.~\cite{Melnikov:2017pgf,Lindert:2017pky,Lindert:2018iug,Neumann:2018bsx,Caola:2018zye,Jones:2018hbb}.

The state-of-the-art for the QCD corrections to differential
distributions in DY production is at a similar level of accuracy. The
total cross section is known fully differentially in the Born phase
space up to
NNLO~\cite{Hamberg:1990np,vanNeerven:1991gh,Anastasiou:2003yy,Melnikov:2006di,Melnikov:2006kv,Catani:2010en,Catani:2009sm,Gavin:2010az,Anastasiou:2003ds},
while differential distributions in transverse momentum were recently
computed up to NNLO both for
$Z$-~\cite{Ridder:2015dxa,Ridder:2016nkl,Gehrmann-DeRidder:2016jns,Gauld:2017tww,Boughezal:2015ded,Boughezal:2016isb}
and
$W$-boson~\cite{Boughezal:2015dva,Boughezal:2016dtm,Gehrmann-DeRidder:2017mvr}
production.  In the DY distributions, electroweak corrections become
important especially at large transverse momenta, and they have been
computed to NLO
in~\cite{Kuhn:2005az,Kuhn:2007qc,Denner:2009gj,Denner:2011vu}.

Although fixed-order results are crucial to obtain reliable
theoretical predictions away from the soft and collinear regions of
the phase space ($v\sim 1$), it is well known that 
regions dominated by soft and collinear QCD radiation---which give rise to the
bulk of the total cross section---are affected by
large logarithmic terms of the form $\as^n \ln^k(1/v)/v$, with
$k\leq 2n-1$, which spoil the convergence of the perturbative series
at small $v$.
In order to have a finite and well-behaved calculation in this limit, the subtraction
of the infrared and collinear divergences requires an all-order
resummation of the logarithmically divergent terms. The logarithmic
accuracy is commonly defined in terms of the perturbative series of the
\emph{logarithm} of the cumulative cross section $\Sigma$ as
\begin{align}
\label{eq:cumulant-initial}
  \ln \Sigma(v) &\equiv \ln\int_0^v \rd v' \; \frac{\rd \Sigma(v')}{\rd v'} \notag\\
&= \sum_n \left\{{\cal O}\left(\as^n\ln^{n+1}(1/v)\right) + {\cal O}\left(\as^n\ln^{n}(1/v)\right) + {\cal O}\left(\as^n\ln^{n-1}(1/v)\right)+\dots\right\}.
\end{align}
One refers to the dominant terms $\as^n \ln^{n+1}(1/v)$ as
leading logarithmic (LL), to terms $\as^n
\ln^{n}(1/v)$ as next-to-leading logarithmic (NLL), to
$\as^n \ln^{n-1}(1/v)$ as next-to-next-to-leading logarithmic
(NNLL), and so on.

The resummation of the $p_t$ spectrum of a heavy colour singlet is
commonly performed in impact-parameter ($b$)
space~\cite{Parisi:1979se,Collins:1984kg}, where the observable
completely factorises and the resummed cross section takes an exponential form.
Using the $b$-space formulation the Higgs $p_t$ spectrum was resummed
at NNLL accuracy in
Refs.~\cite{Bozzi:2005wk,deFlorian:2012mx,Becher:2012yn}, following
either the conventional approach of Ref.~\cite{Collins:1984kg}, or a
soft-collinear-effective-theory~\cite{Bauer:2000ew,Bauer:2000yr,Bauer:2001yt,Bauer:2002nz}
(SCET) formulation of Refs.~\cite{Becher:2010tm,GarciaEchevarria:2011rb}. A
study of the related theory uncertainties in the SCET formulation was
presented in Ref.~\cite{Neill:2015roa}.  In DY production, NNLL
predictions for the transverse momentum of the color singlet as well
as for $\phs$ were obtained in
Refs.~\cite{Bozzi:2010xn,Becher:2010tm,Banfi:2012du}.  The impact of
both threshold and high-energy resummation on the
small-transverse-momentum region was also studied in detail in
Refs.~\cite{Li:1998is,Laenen:2000ij,Kulesza:2003wn,Marzani:2015oyb,Forte:2015gve,Caola:2016upw,Lustermans:2016nvk,Marzani:2016smx,Muselli:2017bad} and the effects were found to be quite moderate at LHC energies.

The problem of the resummation of the transverse-momentum distribution
in direct ($p_t$) space received substantial attention throughout the
years~\cite{Ellis:1997ii,Frixione:1998dw,Kulesza:1999sg}, but remained
unsolved until recently. Due to the vectorial nature of $p_t$
(analogous considerations apply to $\phs$), it is indeed not possible
to define a resummed cross section at a given logarithmic accuracy in
direct space that is simultaneously free of both subleading-logarithmic
contributions and spurious singularities at finite, non-zero values
of $p_t$. A possible solution to the problem was recently proposed in
Refs.~\cite{Monni:2016ktx,Bizon:2017rah}, in whose formalism the
resummation is performed by generating the relevant QCD radiation by
means of a Monte Carlo (MC) algorithm.
The resummation of the $p_t$ spectrum in momentum space has been also
studied in Ref.~\cite{Ebert:2016gcn} within a SCET framework, where the
renormalisation-group evolution is performed directly in $p_t$
space. An alternative technique to analytically transform the
impact-parameter-space result into momentum space was recently
proposed in Ref.~\cite{Kang:2017cjk}.

All the necessary ingredients for the N$^3$LL resummation of $p_t$
(and $\phs$) spectra in color-singlet production have been computed
in~\cite{Catani:2011kr,Catani:2012qa,Gehrmann:2014yya,Echevarria:2016scs,Li:2016ctv,Vladimirov:2016dll},
and the four-loop cusp anomalous dimension has been recently obtained
numerically in refs.~\cite{Moch:2017uml,Moch:2018wjh}.  This has paved
the way to more accurate theoretical results for transverse
observables in the infrared region, like for instance the computation
of the Higgs-transverse-momentum spectrum at N$^3$LL matched to NNLO
in Refs.~\cite{Bizon:2017rah,Chen:2018pzu}.  In this manuscript,
employing the direct-space resummation at N$^3$LL accuracy of
Ref.~\cite{Bizon:2017rah} matched to NNLO, we present results for
Higgs $p_t$ both at the inclusive level and with fiducial cuts on the
decay products in the $H\to \gamma\gamma$ channel.  We also consider
Drell-Yan pair production and compute N$^3$LL+NNLO predictions for the
transverse momentum of the lepton pair and for the $\phs$ observable,
comparing these results to ATLAS measurements at $8~\TeV$.

The article is organised as follows. In section~\ref{sec:fixedorder}
we discuss the computation of the NNLO differential distributions in
DY and H production with the fixed-order parton-level code \nnlojet.
Section~\ref{sec:resummation} contains a brief review of the
resummation for the $p_t$ and $\phs$ distributions using a
momentum-space approach as implemented in the computer code {\tt
  RadISH}, and in section~\ref{sec:matching} we discuss in detail the
matching to fixed order together with the validation of our
calculation. Section~\ref{sec:Higgs} reports the results for H
production, while the analogous results for DY production are reported
in Section~\ref{sec:DY}. Section~\ref{sec:conclusions} contains our
conclusions. We report the relevant formulae used for the matching in
Appendix \ref{app:matching}, while Appendix \ref{app:sudakov-radiator}
contains various quantities necessary for the resummation up to
N$^3$LL.

\section{Fixed order}
\label{sec:fixedorder}
\noindent

In this article we consider the production of either a Higgs boson or
a leptonic Drell-Yan pair.  In particular, the main focus lies in the
description of the transverse-momentum spectrum and, in the case of DY
production, of the closely related $\phs$ observable.  These
observables are studied in the context of matching the fixed-order
calculation to a resummed prediction, and consequently the low- to
intermediate-$p_t$ regimes are of particular interest.

For the Higgs production process, we therefore restrict ourselves to the region with $\pth \lesssim m_t$ where the HEFT description is appropriate.
In this effective-field-theory framework, the top quark is integrated out in the large-top-mass limit ($m_t\rightarrow\infty$), giving rise to an effective operator that directly couples the Higgs field to the gluon field-strength tensor via~\cite{Wilczek:1977zn,Shifman:1978zn,Inami:1982xt}
\begin{equation} 
  \mathcal{L}_\text{HEFT} =
  -\frac{\lambda}{4} \; G^{\mu\nu}G_{\mu\nu}  H .
\end{equation} 
The Wilson coefficient $\lambda$ is known to three-loop accuracy~\cite{Chetyrkin:1997un} and its renormalisation-scale dependence was studied in~\cite{Chen:2016zka}. 
We consider the $\pth$ spectrum for both the inclusive production of an on-shell Higgs boson as well as including its decay into two photons.
For the latter, the production and decay are treated in the narrow-width approximation and fiducial cuts, summarised in Section~\ref{sec:Higgs}, are applied on the photons in the final state.

For the DY process, we consider the full off-shell production of a charged lepton pair, including both the $Z$-boson and photon exchange contributions.
Fiducial cuts are applied to the leptons in the final state and match the corresponding measurement performed by ATLAS at $8~\TeV$~\cite{Aad:2015auj}, which are summarised in Section~\ref{sec:DY}.
We consider both the $\ptz$ spectrum as well as the $\phs$ distribution, which are further studied multi-differentially for different invariant-mass ($\mll$) or rapidity ($\yll$) bins.
\\

The differential distributions in $v=p_t/M,~\phs$ for the production
of a colour singlet at hadron colliders are indirectly generated
through the recoil of the colour singlet against QCD radiation.  The
observables $v$ are therefore closely related to the $X+\jet$ process
with $X=H,~Z$, where the jet requirement is replaced by a restriction
on $v$ to be non-vanishing: $v \geq v_\text{cut} > 0$.  The
state-of-the-art fixed-order QCD predictions for this class of
processes is at
NNLO~\cite{Boughezal:2015dra,Boughezal:2015aha,Caola:2015wna,Chen:2016zka,Ridder:2015dxa,Ridder:2016nkl,Gehrmann-DeRidder:2016jns,Gauld:2017tww,Boughezal:2015ded,Boughezal:2016isb}.
Starting from the LO distributions, in which the colour singlet
recoils against a single parton, the NNLO predictions receive
contributions from configurations (with respect to LO) with two extra
partons (RR: double-real corrections for
H~\cite{DelDuca:2004wt,Dixon:2004za,Badger:2004ty} and
DY~\cite{Campbell:2002tg,Campbell:2003hd,Hagiwara:1988pp,Berends:1988yn,Falck:1989uz}),
with one extra parton and one extra loop (RV: real-virtual corrections
for H~\cite{Dixon:2009uk,Badger:2009hw,Badger:2009vh} and
DY~\cite{Campbell:2002tg,Campbell:2003hd,Glover:1996eh,Bern:1996ka,Campbell:1997tv,Bern:1997sc})
and with no extra parton but two extra loops (VV: double-virtual
corrections for H~\cite{Gehrmann:2011aa} and
DY~\cite{Moch:2002hm,Garland:2001tf,Garland:2002ak,Gehrmann:2011ab}).
Each of the three contributions is separately infrared divergent
either in an implicit manner from phase-space regions where parton
radiations become unresolved (soft and/or collinear) or in a explicit
manner from divergent poles in virtual loop corrections.  Only the sum
of the three contributions is finite.

Our calculation is performed using the parton-level event generator \nnlojet, which implements the antenna subtraction method~\cite{GehrmannDeRidder:2005hi,Daleo:2006xa,Currie:2013vh} to isolate infrared singularities and to enable their cancellation between different contributions prior to the numerical phase-space integration.  
The NNLO corrections for Higgs and DY production at finite $v$ are calculated using established implementations for $pp\rightarrow H+\jet$~\cite{Chen:2014gva,Chen:2016zka} and $pp\rightarrow Z+\jet$~\cite{Ridder:2015dxa,Ridder:2016nkl,Gehrmann-DeRidder:2016jns,Gauld:2017tww} at NNLO, and it takes the schematic form: 
\begin{align}
\label{eq:NNLOwithansub} 
\sigma^\text{NNLO}_{X+\jet}=&\int_{\Phi_{X+3}}\Big(\rd\sigma^{RR}_\text{NNLO}-\rd\sigma^S_\text{NNLO}\Big)\nonumber\\
+&\int_{\Phi_{X+2}}\Big(\rd\sigma^{RV}_\text{NNLO}-\rd\sigma^T_\text{NNLO}\Big)\nonumber\\
+&\int_{\Phi_{X+1}}\Big(\rd\sigma^{VV}_\text{NNLO}-\rd\sigma^U_\text{NNLO}\Big) .
\end{align} 
The antenna subtraction terms, $\rd\sigma^{S,T,U}_\text{NNLO}$, for both Higgs and Drell-Yan related processes are constructed from antenna functions~\cite{GehrmannDeRidder:2005cm,GehrmannDeRidder:2005aw,Daleo:2009yj,Boughezal:2010mc,Gehrmann:2011wi,GehrmannDeRidder:2012ja} to cancel infrared singularities between the contributions of different parton multiplicities. 
The integrals are performed over the phase space ${\Phi_{X+1,2,3}}$ corresponding to the production of the colour singlet in association with one, two or three partons in the final state. 
The integration of the final-state phase space is fully differential such that any infrared-safe observable $\cal O$ can be studied through differential distributions as $\rd\sigma^\text{NNLO}_{X+\jet}/\rd\cal O$.

For large values of $v$ ($v\sim 1$), the phase-space integral in each
line of Eq.~\eqref{eq:NNLOwithansub} is well defined and was
calculated with high numerical precision in previous studies.
Extending these predictions to smaller, but finite $v\ (\sim 0.01)$
becomes extremely challenging due to the wider dynamical range that is
probed in the integration.  Both the matrix elements and the
subtraction terms grow rapidly in magnitude towards smaller values of
$v$, thereby resulting in large numerical cancellations between them
and rendering both the numerical precision and the stability of the
results challenging.  The finite remainder of such cancellations needs
to be numerically stable in order to be consistently combined with
resummed logarithmic corrections and extrapolated to the limit
$v\to 0$.  For this reason, the integration is performed separately
for each individual initial-state partonic channel. We further split
the integration region for each channel into multiple intervals in
$v$, which are partially overlapping with each other.  By carefully
checking the consistency of the distributions in the overlapping
region and using dedicated reweighting factors in each interval, we
use \nnlojet to produce fixed-order predictions up to NNLO for values
in $v$ down to $p_t=2~\GeV$ and
$\phs = 0.004$~\cite{Gehrmann-DeRidder:2016jns}.

The accuracy of the results obtained with the \nnlojet code for small
$v$ has been systematically validated in Ref.~\cite{Chen:2018pzu} by
comparing fixed-order predictions of the Higgs boson transverse
momentum distribution $\rd\sigma_\text{NNLO}/\rd \pth$ against the
expansion of the N$^3$LL resummation (obtained in the framework of
soft-collinear effective field theory, SCET) to the respective order
in the small $\pth$ region.  This validation was performed for
individual initial-state partonic channels down to $\pth=0.7~\GeV$.

As $v\rightarrow 0$, the final-state phase space $\Phi_{X+1,2,3}$ is
reduced to the phase space of colour singlet production $\Phi_{X}$.
The RR, RV, and VV contributions contain infrared divergences with one
extra unresolved parton that cannot be cancelled by the subtraction
terms $\rd\sigma^{S,T,U}_\text{NNLO}$.  These extra logarithmic
divergences are cancelled by combining the fixed-order computation to
a resummed calculation, where the logarithms in the fixed-order
prediction are subtracted and replaced by a summation of the
corresponding enhanced terms to all orders in perturbation theory.
This operation is discussed in the next section, and more details on
the combination of the two results are reported in
Appendix~\ref{app:matching}.

\section{Resummation}
\label{sec:resummation}
\noindent

The approach developed in Refs.~\cite{Monni:2016ktx,Bizon:2017rah}
uses the factorisation properties of the QCD squared amplitudes to
devise a Monte Carlo formulation of the all-order calculation. In this
framework, large logarithms are resummed directly in momentum space by
effectively generating soft and/or collinear emissions in a fashion
similar in spirit to an event generator.

To summarise the final result, we consider the cumulative distribution
\begin{equation}
\label{eq:cumulative}
\Sigma(v) \equiv \int_0^v \rd v'\; \frac{\rd \Sigma(v')}{\rd v'}
\end{equation}
for an observable $v^{(\prime)}=V(\Phi_B,k_1,\dots,k_n)$, being either
$p_t/M$ or $\phs$, in the presence of $n$ real emissions with momenta
$k_1,...,k_n$. Using the notation of Ref.~\cite{Bizon:2017rah},
$\Sigma(v)$ can be expressed as
\begin{equation}
  \label{eq:Sigma-2}
  \Sigma(v) = \int \rd\Phi_B {\cal V}(\Phi_B) \sum_{n=0}^{\infty}
  \int\prod_{i=1}^n [\rd k_i]
  |{\cal M}(\Phi_B,k_1,\dots ,k_n)|^2\,\Theta\left(v-V(\Phi_B,k_1,\dots,k_n)\right)\,,
\end{equation}
where ${\cal M}$ is the matrix element for $n$ real emissions and
${\cal V}(\Phi_B)$ denotes the resummed form factor that encodes the
purely virtual corrections~\cite{Dixon:2008gr}.  The phase spaces of
the $i$-th emission $k_i$ and that of the Born
configuration%
\footnote{In the context of resummation, the Born
  configuration denotes the production of the colour-singlet state
  without any extra radiation. This should not be confused with the
  fixed-order counting of orders, where LO denotes the production of
  the colour-singlet state recoiling against a parton at finite
  transverse momentum.} are denoted by $[\rd k_i]$ and $\rd\Phi_B$,
respectively.

The recursive infrared and collinear (rIRC) safety~\cite{Banfi:2004yd}
of the observable allows one to establish a well defined logarithmic
counting in the squared amplitude~\cite{Banfi:2004yd,Banfi:2014sua},
and to systematically identify the contributions that enter at a
given logarithmic order.
In particular, the squared amplitude can be decomposed in terms of
$n$-particle-correlated blocks, such that blocks with $n$ particles start
contributing one logarithmic order higher than blocks with $n-1$ particles. 

Eq.~\eqref{eq:Sigma-2} contains exponentiated divergences of virtual origin
in the $\mathcal V(\Phi_B) $ factor, as well as singularities in the real
matrix elements, which appear at all perturbative orders.
In order to handle such divergences, one can introduce a resolution
scale $Q_0$ on the transverse momentum of the radiation:
thanks to rIRC safety, {\it unresolved} real radiation (i.e.~softer
than $Q_0$) does not contribute to the observable's value, namely it
can be neglected when computing $V(\Phi_B,k_1,\dots,k_n)$, thus it
exponentiates and cancels the divergences contained in
$\mathcal V(\Phi_B)$ at all orders. The precise definition of the
unresolved radiation requires a careful clustering of momenta
belonging to a given correlated block in order to be collinear
safe. On the other hand, the real radiation harder than the resolution
scale (referred to as {\it resolved}) must be generated exclusively
since it is constrained by the $\Theta$ function in
Eq.~\eqref{eq:Sigma-2}. rIRC safety also ensures that the dependence
of the results upon $Q_0$ is power-like, hence the limit $Q_0\to0$ can
be taken safely.\\

For observables which depend on the total transverse momentum of QCD
radiation, such as $p_t$ or $\phs$, it is particularly convenient to
set the resolution scale to a small fraction $\eps > 0$ of the
transverse momentum of the block with largest $k_t$, hereby denoted
by $\eps k_{t1}$, which allows for an efficient Monte Carlo
implementation of the resulting resummed formula that can be used
to simultaneously compute both $p_t$ and $\phs$.

\noindent Including terms up to N$^3$LL, the cumulative cross section
in momentum space can be recast in the following
form~\cite{Bizon:2017rah}%
\footnote{We have split the result into a
  sum of three terms. The first term contains the full NLL
  corrections. The second term of Eq.~\eqref{eq:master-kt-space}
  (first set of curly brackets) starts contributing at NNLL accuracy,
  while the third term (second set of curly brackets) is purely
  N$^3$LL.}
\begin{align}
\label{eq:master-kt-space}
&\frac{\rd\Sigma(v)}{\rd\Phi_B} =\int_0^\infty\frac{\rd k_{t1}}{k_{t1}}{\cal J}(k_{t1})\frac{\rd
  \phi_1}{2\pi}\partial_{\tilde L}\left(-\re^{-\tilde R(k_{t1})} {\tilde{\cal L}}_{\rm
  N^3LL}(k_{t1}) \right) \int \dZ\Theta\left(v-V(\Phi_B,k_1,\dots, k_{n+1})\right)
                             \notag\\\notag\\
& + \int_0^\infty\frac{\rd k_{t1}}{k_{t1}}{\cal J}(k_{t1})\frac{\rd
  \phi_1}{2\pi} \re^{-\tilde R(k_{t1})} \int \dZ\int_{0}^{1}\frac{\rd \zeta_{s}}{\zeta_{s}}\frac{\rd
  \phi_s}{2\pi}\Bigg\{\bigg({\tilde R}' (k_{t1}) {\tilde{\cal L}}_{\rm
  NNLL}(k_{t1}) - \partial_{\tilde L} {\tilde{\cal L}}_{\rm
  NNLL}(k_{t1})\bigg)\notag\\
&\times\left({\tilde R}'' (k_{t1})\ln\frac{1}{\zeta_s} +\frac{1}{2} {\tilde R}'''
  (k_{t1})\ln^2\frac{1}{\zeta_s} \right) - {\tilde R}' (k_{t1})\left(\partial_{\tilde L} {\tilde{\cal L}}_{\rm
  NNLL}(k_{t1}) - 2\frac{\beta_0}{\pi}\as^2(k_{t1}) \hat{P}^{(0)}\otimes {\tilde{\cal L}}_{\rm
  NLL}(k_{t1}) \ln\frac{1}{\zeta_s}
\right)\notag\\
&+\frac{\as^2(k_{t1}) }{\pi^2}\hat{P}^{(0)}\otimes \hat{P}^{(0)}\otimes {\tilde{\cal L}}_{\rm
  NLL}(k_{t1})\Bigg\} \bigg\{\Theta\left(v-V(\Phi_B,k_1,\dots,
  k_{n+1},k_s)\right) - \Theta\left(v-V(\Phi_B,k_1,\dots,
  k_{n+1})\right)\bigg\}\notag\\\notag\\
& + \frac{1}{2}\int_0^\infty\frac{\rd k_{t1}}{k_{t1}}{\cal J}(k_{t1})\frac{\rd
  \phi_1}{2\pi} \re^{-\tilde R(k_{t1})} \int \dZ\int_{0}^{1}\frac{\rd \zeta_{s1}}{\zeta_{s1}}\frac{\rd
  \phi_{s1}}{2\pi}\int_{0}^{1}\frac{\rd \zeta_{s2}}{\zeta_{s2}}\frac{\rd
  \phi_{s2}}{2\pi} {\tilde R}' (k_{t1})\notag\\
&\times\Bigg\{ {\tilde{\cal L}}_{\rm
  NLL}(k_{t1}) \left({\tilde R}'' (k_{t1})\right)^2\ln\frac{1}{\zeta_{s1}} \ln\frac{1}{\zeta_{s2}} - \partial_{\tilde L} {\tilde{\cal L}}_{\rm
  NLL}(k_{t1}) {\tilde R}'' (k_{t1})\bigg(\ln\frac{1}{\zeta_{s1}}
  +\ln\frac{1}{\zeta_{s2}} \bigg)\notag\\
&+ \frac{\as^2(k_{t1}) }{\pi^2}\hat{P}^{(0)}\otimes \hat{P}^{(0)}\otimes {\tilde{\cal L}}_{\rm
  NLL}(k_{t1})\Bigg\}\notag\\
&\times \bigg\{\Theta\left(v-V(\Phi_B,k_1,\dots,
  k_{n+1},k_{s1},k_{s2})\right) - \Theta\left(v-V(\Phi_B,k_1,\dots,
  k_{n+1},k_{s1})\right) -\notag\\ &\Theta\left(v-V(\Phi_B,k_1,\dots,
  k_{n+1},k_{s2})\right) + \Theta\left(v-V(\Phi_B,k_1,\dots,
  k_{n+1})\right)\bigg\} + {\cal O}\left(\as^n \ln^{2n -
                                    6}\frac{1}{v}\right),
\end{align}
where $\zeta_{si} \equiv k_{tsi}/k_{t1}$ and we introduced the notation $\dZ$ to denote an ensemble that
describes the emission of $n$ identical independent blocks~\cite{Bizon:2017rah}.
The average of a function $G(\{\tilde p\},\{k_i\})$ over the measure
$\rd {\cal Z}$ is defined as ($\zeta_{i} \equiv k_{ti}/k_{t1}$)
\begin{equation}
\label{eq:dZ}
\begin{split}
\int \dZ  G(\{\tilde p\},\{k_i\})=\re^{-{\tilde R}'(k_{t1})\ln\frac{1}{\eps}}
   \sum_{n=0}^{\infty}\frac{1}{n!} \prod_{i=2}^{n+1}
    \int_{\eps}^{1} \frac{\rd\zeta_i}{\zeta_i}\int_0^{2\pi}
   \frac{\rd\phi_i}{2\pi} {\tilde R}'(k_{t1})G(\{\tilde p\},k_1,\dots,k_{n+1})\,.
\end{split}
\end{equation}
The $\ln 1/\eps$ divergence in the exponential prefactor of
Eq.~\eqref{eq:dZ} cancels exactly against that contained in the
resolved real radiation, encoded in the nested sums of products on the
right-hand side of the same equation. This ensures that the final
result is therefore $\eps$-independent.

To obtain Eq.~\eqref{eq:master-kt-space} we used the fact that, for
resolved radiation, $\zeta_i$ is a quantity of
$\mathcal O(1)$, which allows us to expand all ingredients in
Eq.~\eqref{eq:master-kt-space} about $k_{t1}$, retaining only terms
necessary for the desired logarithmic accuracy. We stress that this is
allowed because of rIRC safety, which ensures that blocks with
$k_{ti} \ll k_{t1}$ do not contribute to the value of the observable and
are therefore fully cancelled by the term 
$\exp\{-{\tilde R}'(k_{t1})\ln(1/\eps)\}$ of Eq.~\eqref{eq:dZ}. Although
not strictly necessary, this expansion allows for a more efficient
numerical implementation. The expansion gives rise to the terms
${\tilde R}^{(n)}$ which denote the derivatives of the radiator as
\begin{equation}
{\tilde R}'= \rd \tilde{R}/\rd\tilde{L},\qquad {\tilde R}''= \rd
      {\tilde R}'/\rd\tilde{L},\qquad {\tilde R}'''= \rd {\tilde R}''/\rd\tilde{L},
\end{equation}
where $\tilde{R}$ takes the form
\begin{align}
\label{eq:mod-radiator}
\tilde{R}(k_{t1}) &= - \tilde{L} g_1(\as \beta_0\tilde{L} ) -
  g_2(\as \beta_0\tilde{L} ) - \frac{\as}{\pi}
  g_3(\as \beta_0\tilde{L} ) - \frac{\as^2}{\pi^2}
  g_4(\as \beta_0\tilde{L} ),
\end{align}
and $\as = \as(\mu_R)$. We report the functions $g_i$ in
Appendix~\ref{app:sudakov-radiator}, and we refer to
Ref.~\cite{Bizon:2017rah} for further details. 
The function $g_4$ involves a contribution from the recently
determined~\cite{Moch:2018wjh} four-loop cusp anomalous dimension
$\Gamma_{\rm cusp}^{(4)}$ that we report in Eq.~\eqref{eq:gamma4}. 

In previous N$^3$LL resummation studies, $\Gamma_{\rm cusp}^{(4)}$ was
either neglected~\cite{Bizon:2017rah,Chen:2018pzu} or extrapolated
from its lower order contributions through a Pad\'e
approximation~\cite{Moch:2005ba}. With the new result
of~\cite{Moch:2018wjh} at hand, we could now explicitly verify that
the numerical impact of $\Gamma_{\rm cusp}^{(4)}$ is indeed very small
(not visibly noticeable in the distributions), and well below other
sources of parametric uncertainties that are discussed in the
following.

The expression in Eq.~\eqref{eq:master-kt-space} would originally
contain resummed logarithms of the form $\ln(Q/k_{t1})$, where $Q$ is
the resummation scale, whose variation is used to probe the size of
subleading logarithmic corrections not included in our result.
In order to ensure that the resummation does not affect the hard
region of the spectrum when matched to fixed order (see
Section~\ref{sec:matching}), the resummed logarithms are supplemented
with power-suppressed terms, negligible at small $k_{t1}$, that ensure
resummation effects to vanish for $k_{t1}\gg Q$. Such {\it modified}
logarithms $\tilde L$ are defined by constraining the rapidity
integration of the real radiation to vanish at large transverse
momenta. This is done by mapping the limit $k_{t1}\to Q$
onto $k_{t1}\to \infty$ in all terms of
Eq.~\eqref{eq:master-kt-space}, with the exception of the observable's
measurement function. A convenient choice of such a mapping is
\begin{equation}
\label{eq:modified-log}
\ln\frac{Q}{k_{t1}} \to \tilde{L}=
\frac{1}{p}\ln\left(\left(\frac{Q}{k_{t1}}\right)^{p} + 1\right),
\end{equation}
where $p$ is a positive real parameter chosen in such a way that the
resummed differential distribution vanishes faster than the fixed-order
one at large $v$, with slope $(1/v)^{p+1}$.
The above prescription comes with the prefactor
$\mathcal J$, defined as
\begin{equation}
\label{eq:jakobian}
{\cal J}(k_{t1}) = \left(\frac{Q}{k_{t1}} \right)^p \left(1+\left(\frac{Q}{k_{t1}} \right)^p\right)^{-1}.
\end{equation}
This corresponds to the Jacobian for the
transformation~\eqref{eq:modified-log}, and ensures the absence of
fractional (although power suppressed) $\as$ powers in the final
distribution~\cite{Bizon:2017rah}.  This factor, once again, leaves
the small $k_{t1}$ region untouched, and only modifies the large
$p_t$ region by power-suppressed effects. Although this
procedure seems a simple change of variables, we stress that the
observable's measurement function (i.e. the $\Theta$ function in
Eq.~\eqref{eq:master-kt-space}) is not affected by this
prescription. As a consequence, the final result will depend on the
parameter $p$ through power-suppressed terms.

The factors ${\tilde {\cal L}}$ contain the parton luminosities up to
N$^3$LL, multiplied by the Born-level squared, and virtual
amplitudes. They are defined as (we adopt the notation of
  Ref.~\cite{Bizon:2017rah})
\begin{align} 
\label{eq:luminosity-NLL}
\tilde{\cal L}_{\rm NLL}(k_{t1}) = \sum_{c, c'}\frac{\rd|\mathcal{M}_{B}|_{cc'}^2}{\rd\Phi_B} f_c\!\left(\mu_F \re^{-\tilde{L}},x_1\right)f_{c'}\!\left(\mu_F \re^{-\tilde{L}},x_2\right),
\end{align}
\begin{align}
\label{eq:luminosity-NNLL}
&\tilde{\cal L}_{\rm NNLL}(k_{t1}) = \sum_{c, c'}\frac{\rd|\mathcal{M}_{B}|_{cc'}^2}{\rd\Phi_B} \sum_{i, j}\int_{x_1}^{1}\frac{\rd z_1}{z_1}\int_{x_2}^{1}\frac{\rd z_2}{z_2}f_i\!\left(\mu_F \re^{-\tilde{L}},\frac{x_1}{z_1}\right)f_{j}\!\left(\mu_F \re^{-\tilde{L}},\frac{x_2}{z_2}\right)\notag\\&\times\Bigg\{\delta_{ci}\delta_{c'j}\delta(1-z_1)\delta(1-z_2)
\left(1+\frac{\as(\mu_R)}{2\pi} \tilde{H}^{(1)}(\mu_R,x_Q)\right) \notag\\
&+ \frac{\as(\mu_R)}{2\pi}\frac{1}{1-2\as(\mu_R)\beta_0
  \tilde{L}}\left(\tilde{C}_{c i}^{(1)}(z_1,\mu_F,x_Q)\delta(1-z_2)\delta_{c'j}+
  \{z_1\leftrightarrow z_2; c,i \leftrightarrow c'j\}\right)\Bigg\},
\end{align}
\begin{align}
\label{eq:mod-luminosity-N3LL}
&\tilde{\cal L}_{\rm N^3LL}(k_{t1})=\sum_{c,
  c'}\frac{\rd|\mathcal{M}_{B}|_{cc'}^2}{\rd\Phi_B} \sum_{i, j}\int_{x_1}^{1}\frac{\rd
  z_1}{z_1}\int_{x_2}^{1}\frac{\rd z_2}{z_2}f_i\!\left(\mu_F \re^{-\tilde{L}},\frac{x_1}{z_1}\right)f_{j}\!\left(\mu_F \re^{-\tilde{L}},\frac{x_2}{z_2}\right)\notag\\&\times\Bigg\{\delta_{ci}\delta_{c'j}\delta(1-z_1)\delta(1-z_2)
\left(1+\frac{\as(\mu_R)}{2\pi} \tilde{H}^{(1)}(\mu_R,x_Q) + \frac{\as^2(\mu_R)}{(2\pi)^2} \tilde{H}^{(2)}(\mu_R,x_Q)\right) \notag\\
&+ \frac{\as(\mu_R)}{2\pi}\frac{1}{1-2\as(\mu_R)\beta_0 \tilde{L}}\left(1- \as(\mu_R)\frac{\beta_1}{\beta_0}\frac{\ln\left(1-2\as(\mu_R)\beta_0 \tilde{L}\right)}{1-2\as(\mu_R)\beta_0 \tilde{L}}\right)\notag\\
&\times\left(\tilde{C}_{c i}^{(1)}(z_1,\mu_F,x_Q)\delta(1-z_2)\delta_{c'j}+ \{z_1\leftrightarrow z_2; c,i \leftrightarrow c',j\}\right)\notag\\
& +
  \frac{\as^2(\mu_R)}{(2\pi)^2}\frac{1}{(1-2\as(\mu_R)\beta_0
  \tilde{L})^2}\Bigg(\tilde{C}_{c i}^{(2)}(z_1,\mu_F,x_Q)\delta(1-z_2)\delta_{c'j} + \{z_1\leftrightarrow z_2; c,i \leftrightarrow c',j\}\Bigg) \notag\\&+  \frac{\as^2(\mu_R)}{(2\pi)^2}\frac{1}{(1-2\as(\mu_R)\beta_0 \tilde{L})^2}\Big(\tilde{C}_{c i}^{(1)}(z_1,\mu_F,x_Q)\tilde{C}_{c' j}^{(1)}(z_2,\mu_F,x_Q) + G_{c i}^{(1)}(z_1)G_{c' j}^{(1)}(z_2)\Big) \notag\\
& + \frac{\as^2(\mu_R)}{(2\pi)^2} \tilde{H}^{(1)}(\mu_R,x_Q)\frac{1}{1-2\as(\mu_R)\beta_0 \tilde{L}}\Big(\tilde{C}_{c i}^{(1)}(z_1,\mu_F,x_Q)\delta(1-z_2)\delta_{c'j} + \{z_1\leftrightarrow z_2; c,i \leftrightarrow c',j\}\Big) \Bigg\}.
\end{align}
where 
\begin{align}
x_1 &= \frac{M}{\sqrt{s}} \; \re^{Y}, &
x_2 &= \frac{M}{\sqrt{s}} \; \re^{-Y},
\end{align}
$Y$ is the rapidity of the colour singlet in the centre-of-mass
frame of the collision at the Born-level, $|\mathcal{M}_{B}|_{cc'}^2$ is
the Born-level squared matrix element, and $x_Q=Q/M$. The above luminosities contain the
NLO and NNLO coefficient functions $\tilde{C}_{ci}^{(n)}$ for Higgs
and Drell-Yan
production~\cite{Catani:2011kr,Catani:2012qa,Gehrmann:2014yya,Echevarria:2016scs}, as
well as the hard virtual corrections $\tilde{H}^{(n)}$. A precise
definition is given is Section 4 of Ref.~\cite{Bizon:2017rah}, and
the relevant formulae are also reported in Appendix~\ref{app:sudakov-radiator}.

Finally, we define the convolution of a regularised splitting function
${\hat P}$~\cite{Moch:2004pa,Vogt:2004mw} with the coefficient
$\tilde{\cal L}_{\rm NLL}$ as
\begin{align} 
\label{eq:Pluminosity-NLL}
\hat{P}^{(0)}\otimes\tilde{\cal L}_{\rm NLL}(k_{t1}) &\equiv \sum_{c,
  c'}\frac{\rd|\mathcal{M}_{B}|_{cc'}^2}{\rd\Phi_B} \bigg\{\left(\hat{P}^{(0)}\otimes
  f\right)_c\left(\mu_F \re^{-\tilde{L}},x_1\right)f_{c'}\!\left(\mu_F \re^{-\tilde{L}},x_2\right)
  \notag\\
&\hspace{3.5cm}+
  f_c \!\left(\mu_F \re^{-\tilde{L}},x_1\right) \left(\hat{P}^{(0)}\otimes f\right)_{c'}\left(\mu_F \re^{-\tilde{L}},x_2\right) \bigg\}.
\end{align}
The term $\hat{P}^{(0)}\otimes \hat{P}^{(0)}\otimes \tilde{\cal L}_{\rm
  NLL}(k_{t1})$ is to be interpreted similarly as
\begin{align}
\hat{P}^{(0)}\otimes \hat{P}^{(0)}&\otimes\tilde{\cal L}_{\rm NLL}(k_{t1})  \equiv \sum_{c,
  c'}\frac{\rd|M_{B}|_{cc'}^2}{\rd\Phi_B} \bigg\{\left(\hat{P}^{(0)}\otimes \hat{P}^{(0)}\otimes
  f\right)_c\left(\mu_F \re^{-\tilde{L}},x_1\right)f_{c'}\!\left(\mu_F \re^{-\tilde{L}},x_2\right)
  \notag\\
&+
  f_c \!\left(\mu_F \re^{-\tilde{L}},x_1\right) \left(\hat{P}^{(0)}\otimes
  \hat{P}^{(0)}\otimes f\right)_{c'}\left(\mu_F
  \re^{-\tilde{L}},x_2\right)  \notag\\
&+ 2\left(\hat{P}^{(0)}\otimes f\right)_c \!\left(\mu_F \re^{-\tilde{L}},x_1\right) \left(\hat{P}^{(0)}\otimes f\right)_{c'}\left(\mu_F \re^{-\tilde{L}},x_2\right) \bigg\}.
\end{align}
Moreover, the explicit factors of the strong coupling evaluated at
$k_{t1}$ in Eq.~\eqref{eq:master-kt-space} are defined as
\begin{equation}
\label{eq:N3LLcoupling}
\as(k_{t1})\equiv \frac{\as(\mu_R)}{1-2\as(\mu_R)\beta_0
  \tilde{L}}.
\end{equation}

\section{Matching to fixed order}
\label{sec:matching}

In this section we discuss the matching of the resummed and the
fixed-order results. Since we work at the level of the cumulative
distribution $\Sigma$, we define the analogue of
Eq.~\eqref{eq:cumulative} for the fixed-order prediction as
\begin{equation}
\Sigma^{\rm N^3LO}(v) = \sigma_{\rm tot}^{\rm N^3LO} - \int_{v}^{\infty} \rd v' \; \frac{\rd \Sigma^{\rm NNLO}(v')}{\rd v'},
\end{equation}
where $\sigma_{\rm tot}^{\rm N^3LO}$ is the total cross section for
the considered processes and $\rd \Sigma^{\rm NNLO}/\rd v'$ denotes the
NNLO differential distribution.

For inclusive Higgs production, the transverse-momentum distribution
at NNLO was obtained in Refs.~\cite{Boughezal:2015aha,Boughezal:2015dra,Caola:2015wna,Chen:2016zka}, while the N$^3$LO total
cross section has been computed in Refs.~\cite{Anastasiou:2016cez,Mistlberger:2018etf}. On the other hand,
the N$^3$LO cross section within fiducial cuts on the Born kinematics
is currently unknown. Since in this article we address differential
distributions for $H\to\gamma\gamma$ with fiducial cuts, we
approximate the N$^3$LO correction to $\sigma_{\rm tot}^{\rm N^3LO}$
by rescaling the NNLO fiducial cross section by the inclusive
(i.e.~without fiducial cuts) N$^3$LO/NNLO $K$ factor. We stress that,
at the level of the differential distributions we are interested in,
this approximation is formally a N$^4$LL effect, and it lies beyond the accuracy considered in this study.

For DY production, the differential distributions to NNLO were
obtained in Refs.~\cite{Gehrmann-DeRidder:2016jns,Boughezal:2015ded}.
We set to zero the unknown N$^3$LO correction to the total cross section,
observing once again that its contribution to the distributions
derived here is subleading.\\

In order to assess the uncertainty associated with the matching
procedure, we consider here two different matching schemes. The first
scheme we introduce is the common additive scheme
defined as
\begin{equation}
\label{eq:additive}
\Sigma_{\rm add}^{\rm MAT}(v) = \Sigma^{\rm N^3LL}(v) + \Sigma^{\rm N^3LO}(v) - \Sigma^{\rm EXP}(v),
\end{equation}
where $\Sigma^{\rm EXP}$ denotes the expansion of the resummation formula $\Sigma^{\rm N^3LL}$ to N$^3$LO.

The second scheme we consider belongs to the class of multiplicative
schemes similar to those defined in Refs.~\cite{Banfi:2012yh,Banfi:2012jm,Banfi:2015pju}, and it is schematically defined as
\begin{equation}
\label{eq:multiplicative0}
\Sigma_{\rm mult}^{\rm MAT}(v) = \Sigma^{\rm N^3LL}(v)
\left[\frac{\Sigma^{\rm N^3LO}(v)}{\Sigma^{\rm EXP}(v)}\right]_{\rm
  EXPANDED~TO~N^3LO},
\end{equation}
where the quantity in square brackets is expanded to N$^3$LO. The two
schemes~\eqref{eq:additive},~\eqref{eq:multiplicative0} are equivalent
at the perturbative order we are working at, and differ by N$^4$LO and
N$^4$LL terms. The main difference between the two schemes is that in
the multiplicative approach, unlike in the additive one, higher-order
corrections are damped by the resummation factor $\Sigma^{\rm N^3LL}$
at low $v$. Moreover, this damping occurs in the region where the
fixed-order result may be occasionally affected by numerical
instabilities, hence allowing for a stable matched distribution even
with limited statistics for the NNLO component.

One advantage of the multiplicative solution is that the N$^3$LO
constant terms, of formal N$^4$LL accuracy, are automatically
extracted from the fixed order in the matching procedure, whenever the
N$^3$LO total cross section is known.
We recall that Eq.~\eqref{eq:master-kt-space} resums all towers of
$\ln(1/v)$ up to N$^3$LL, defined at the level of the logarithm of
$\Sigma$~\eqref{eq:cumulant-initial}. At this order, one predicts
correctly all logarithmic terms up to, and including,
$\as^n \ln^{2n-5}(1/v)$ in the expanded formula for $\Sigma$, while
terms of order $\as^n \ln^{2n-6}(1/v)$ would be modified by including
N$^4$LL corrections.

The inclusion of constant terms of order ${\cal O}(\as^3)$ relative to
Born level in the resummed formula, of formal N$^4$LL accuracy,
extends the prediction to all terms of order $\as^n\ln^{2n-6}(1/v)$ in
the expanded formula for $\Sigma$.
Indeed these terms, which contain the N$^3$LO collinear coefficient
functions and three-loop virtual corrections, would multiply the
Sudakov $\re^{-\tilde R(k_{t1})}$ in the resummed
formula~\eqref{eq:master-kt-space} starting at N$^4$LL. Since they are
currently unknown analytically, in an additive matching these terms
are simply added to the resummed cumulative result, and disappear at
the level of the differential distribution. On the other hand, in a
multiplicative scheme, they multiply the resummed cross section and
hence correctly include a whole new tower of N$^4$LL terms
$\as^n\ln^{2n-6}(1/v)$ in the expanded formula for the matched
cumulative cross section $\Sigma^{\rm MAT}$.\footnote{Notice that this
  does not imply that the whole class of N$^4$LL terms is included.
  This would instead require all terms of the form
  $\as^n\ln^{n-3}(1/v)$ in $\ln \Sigma$,
  Eq.~\eqref{eq:cumulant-initial}, which would predict correctly all
  terms $\as^n\ln^{2n-6}(1/v)$ and $\as^n\ln^{2n-7}(1/v)$ in the
  expanded $\Sigma$.}  We stress that this, as pointed out above,
requires the knowledge of the N$^3$LO cross section in the considered
fiducial volume. This is currently only known in the case of fully
inclusive Higgs production, whose results are presented in
Section~\ref{sec:inclusive_h}. In the remaining studies of fiducial
distributions, both for Higgs in Section~\ref{sec:fiducial_h}, and for
DY in Section~\ref{sec:DY}, the N$^3$LO cross sections are
approximated, as described at the beginning of this section, and hence
the tower of N$^4$LL terms $\as^n\ln^{2n-6}(1/v)$ in $\Sigma$ is not
fully included.

However, there is a drawback in using Eq.~\eqref{eq:multiplicative0}
as is. Indeed, in the limit ${\tilde L}\to 0$, $\Sigma^{\rm N^3LL}$
tends to the integral of $\tilde{\cal L}_{\rm N^3LL}(\mu_F)$ (defined
in Eq.~\eqref{eq:mod-luminosity-N3LL}) over $\Phi_B$, evaluated at
${\tilde L}=0$.  Therefore, the fixed-order result
$\Sigma^{\rm N^3LO}$ at large $v$ receives a spurious correction of
relative order $\as^4$
\begin{equation}
\Sigma_{\rm mult}^{\rm MAT}(v) \sim \Sigma^{\rm N^3LO}(v)\left(1+{\cal O}(\as^4)\right).
\end{equation}
Despite being formally of higher
order, this effect can be moderately sizeable in processes with large $K$
factors, such as Higgs production. There are different possible
solutions to this problem. In Ref.~\cite{Bizon:2017rah} the resummed component (as
well as the relative expansion) was modified by introducing a damping factor as
\begin{equation}
\Sigma^{\rm N^3LL} \to \left(\Sigma^{\rm N^3LL}\right)^Z,
\end{equation}
where $Z$  is a $v$-dependent exponent that effectively acts as a smoothened $\Theta$ function that
tends to zero at large $v$. This solution, however, introduces new
parameters that control the scaling of the damping factor $Z$ (see
Section 4.2 of Ref.~\cite{Bizon:2017rah} for details). In this article we adopt a
simpler solution, which avoids the introduction of extra parameters in
the matching scheme. To this end, we define the multiplicative matching
scheme by normalising the resummed prefactor to its
asymptotic $\tilde{L}\to 0$ value. This is simply given by the
integral over the Born phase space $\Phi_{B}$ of the $\tilde{L}\to 0$
limit of $\tilde{\cal
  L}_{\rm N^3LL}$ (that we report in
Eq.~\eqref{eq:lumi_asympt})
\begin{equation}
\label{eq:asypt}
\Sigma^{\rm N^3LL}_{\rm asym.} = \int_{\rm with~cuts} \hspace{-1cm}\rd
\Phi_B \quad \left(\lim_{\tilde{L}\to 0}
\tilde{\cal
  L}_{\rm N^3LL}\right),
\end{equation}
where the integration over $\Phi_B$ is performed by taking into
account the phase-space cuts of the experimental analysis.

We thus obtain
\begin{equation}
\label{eq:multiplicative1}
\Sigma_{\rm mult}^{\rm MAT}(v) = \frac{\Sigma^{\rm N^3LL}(v)}{\Sigma^{\rm N^3LL}_{\rm asym.} } \left[\Sigma^{\rm N^3LL}_{\rm asym.} \frac{\Sigma^{\rm N^3LO}(v)}{\Sigma^{\rm EXP}(v)}\right]_{\rm EXPANDED~TO~N^3LO},
\end{equation}
where
\begin{equation}
\Sigma^{\rm N^3LL}(v)\xrightarrow[v \gg Q/M]{} \Sigma^{\rm N^3LL}_{\rm asym.} ,
\end{equation}
and the whole squared bracket in Eq.~\eqref{eq:multiplicative1} is
expanded to N$^3$LO.
This ensures that, in the $v \gg Q/M$ limit,
Eq.~\eqref{eq:multiplicative1} reproduces by construction the
fixed-order result, and no large spurious higher-order corrections
arise in this region. The detailed matching formulae for the two
schemes considered in our analysis are reported in Appendix~\ref{app:matching}.

In
order to estimate the systematic uncertainty associated with the
choice of the matching scheme, a consistent comparison between the two
will be performed in the next section considering inclusive Higgs
production as a case study.

Before we proceed with the results, we stress that in the remainder of
this article we will only focus on differential distributions rather
than on cumulative ones. Therefore, at the level of the spectrum, in
our notation we will drop one order in the fixed-order counting, so
that the derivative of $\Sigma^{\rm N^3LO}$ will be referred to as
a NNLO distribution, and analogously for the lower-order cases.

In the next two subsections we perform some validation studies both
for Higgs (Section~\ref{sec:validationH}) and DY
(Section~\ref{sec:validationDY}) production, where we compare the
fixed-order calculation in the deep infrared regime to the expansion
of the resummed result. Moreover, we discuss the uncertainty
associated with the choice of the matching scheme, and estimate it
through a comparison of the two prescriptions defined above for a case
study.

\subsection{Validation of the expansion and matching uncertainty for
  Higgs production}
\label{sec:validationH}

\begin{figure}[htbp]
  \centering
  \includegraphics[trim={0 -0.2cm 0 0},width=0.45\linewidth]{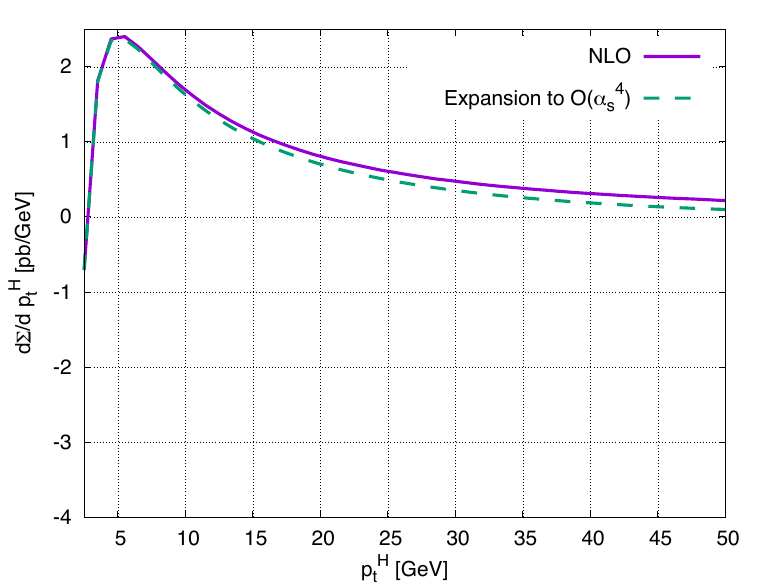} 
  \qquad
  \includegraphics[width=0.45\linewidth]{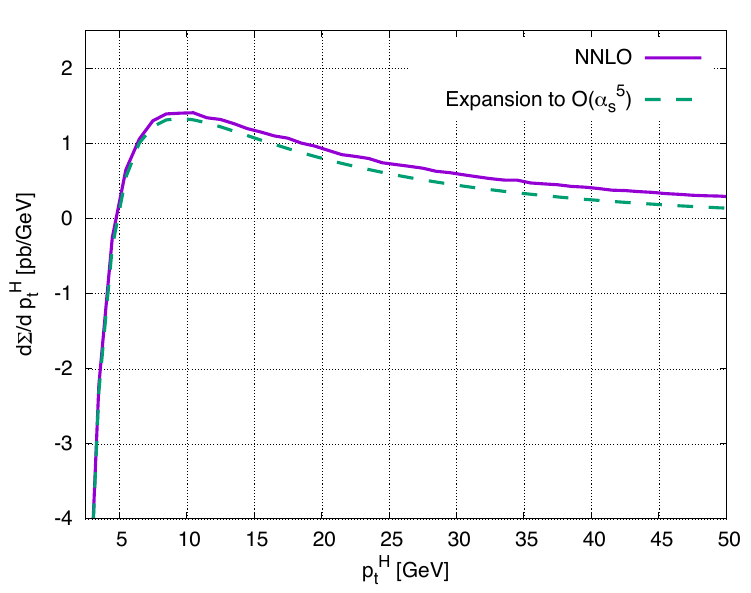} 
  \caption{Comparison between the fixed-order transverse momentum
    distribution for Higgs boson production at $\sqrt{s} = 13~\TeV$ at
    NLO (left) and NNLO (right) and the expansion of the N$^3$LL
    resummation formula given in Eq.~\eqref{eq:master-kt-space} to the
    corresponding order, i.e. ${\cal O}(\as^4)$ and ${\cal O}(\as^5)$
    (namely ${\cal O}(\as^2)$ and ${\cal O}(\as^3)$ relative to Born),
    respectively.}
  \label{fig:res_v_FO}
  %-----
  \bigskip
  \includegraphics[trim={0 -0.6cm 0 0},width=.45\linewidth]{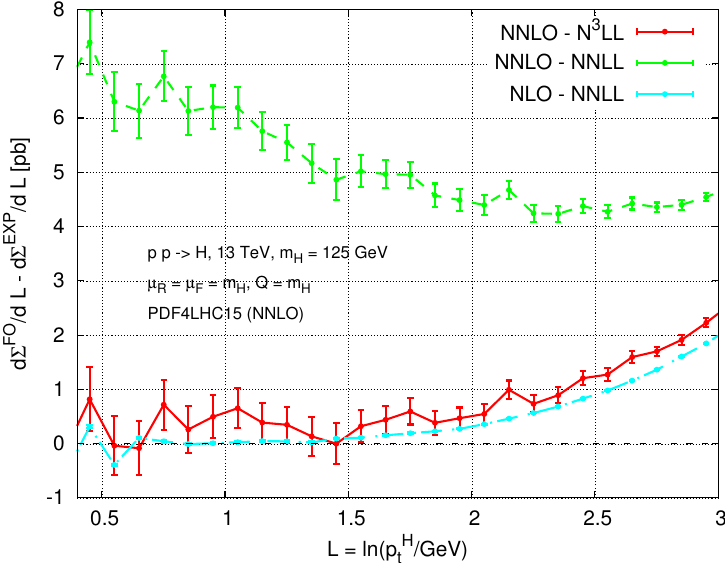}
  \qquad
  \includegraphics[width=.45\linewidth]{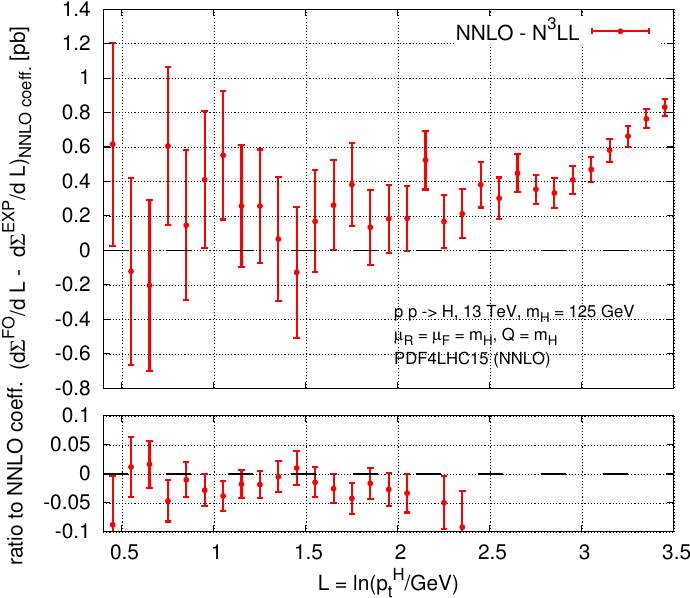} 
  \caption{Left: difference between the full NLO and NNLO $\pth$
    distribution and the expansion of the NNLL and N$^3$LL
    resummation formulae~\eqref{eq:master-kt-space} to the 
    respective perturbative order. 
    Right: difference between the fixed-order NNLO coefficient,
    i.e.\ the $\order{\as^5}$ term alone, 
    and the corresponding coefficient obtained from the expansion of
    the N$^3$LL resummation.}
  \label{fig:res_v_FO_sub}
\end{figure}

To perform the matching to fixed order, the resummation
formula~\eqref{eq:master-kt-space} is expanded up to the third order in
the strong coupling. To obtain the expanded results, one can directly set 
the resolution scale $\eps$ to zero, since the cancellation of IRC
divergences is manifest. In Figure~\ref{fig:res_v_FO} we show the
comparison between the expansion of the N$^3$LL resummed cross section
and the fixed order for the differential distribution of $\pth$ both
at NLO (left plot) and at NNLO (right plot). We remind the reader that
at the level of the differential distribution NNLO denotes the
derivative of the N$^3$LO cumulant, and similarly for lower orders.

In Figure~\ref{fig:res_v_FO} we see that below $\pth\sim 10~\GeV$ the
fixed-order and the expansion of the resummation are in excellent
agreement, and that the size of non-logarithmic terms in the
perturbative series remains moderate up to $\pth\sim 50~\GeV$.

It is instructive to further investigate the difference between
the fixed order and the expansion of the resummation formula in the
region of very small $\pth$. In particular, we consider the differential
distribution
\begin{equation}
\label{eq:logdiff}
\frac{\rd \Sigma(\pth)}{\rd\ln(\pth/\GeV)} \, ,
\end{equation}
in order to highlight potential logarithmic differences in the
$\pth\to0$ region. A similar validation of the NNLO $\pth$
distribution has been performed in Ref.~\cite{Chen:2018pzu}. The
result of our comparison is displayed in the left panel of
Figure~\ref{fig:res_v_FO_sub}. The dashed green line shows the
difference between the NNLO distribution and the ${\cal O}(\as^3)$
expansion of the NNLL resummation. As one expects, at small $\pth$ the
two predictions for the cumulative distribution differ by a
double-logarithmic term (due to the absence of the NNLO coefficient
functions and of the two-loop virtual corrections in the NNLL result),
which induces a linear slope at the level of the differential
distribution~\eqref{eq:logdiff}.  When we include the N$^3$LL
corrections (solid red line), the difference between the two curves
tends to zero, hence proving the consistency between the two
predictions. For comparison, the difference between the NLO and NNLL
(cyan dot-dashed line) is also reported. The right panel of
Figure~\ref{fig:res_v_FO_sub} shows the difference between the NNLO
coefficient and the corresponding expansion of the N$^3$LL resummation
at the same order. The lower inset of the same figure shows the ratio
of the above difference to the NNLO coefficient, which helps quantify
the relative difference.

\begin{figure}[htbp]
  \centering
  \includegraphics[width=.45\linewidth]{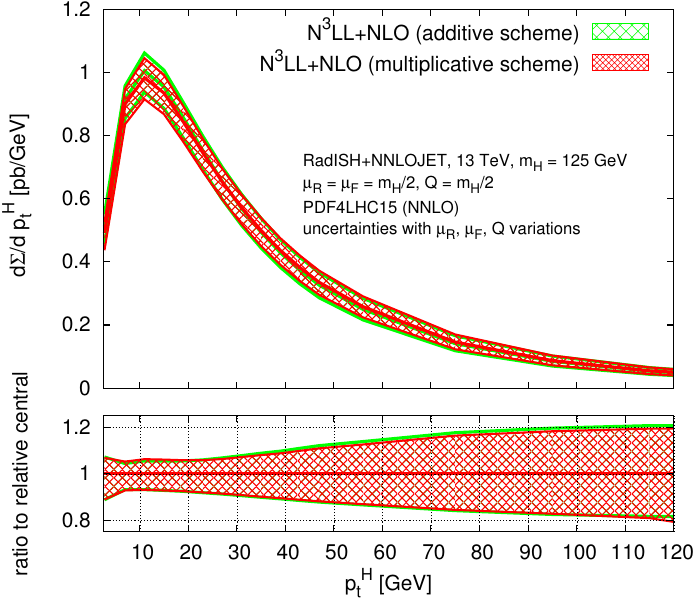} 
  \caption{Comparison between additive and multiplicative matching schemes at N$^3$LL+NLO for the  transverse momentum distribution for Higgs boson production at $\sqrt{s} = 13~\TeV$. The lower panel shows the relative uncertainty bands obtained within the two schemes.}
  \label{fig:scheme_comparison}
\end{figure}

As a check on the theoretical setup that will be used in the next
sections, it is interesting to compare the predictions for the $\pth$
spectrum obtained with the two matching schemes defined in
Eqs.~\eqref{eq:additive} and~\eqref{eq:multiplicative1}. In order to
compare the multiplicative and additive schemes on an equal footing,
hence including the same ingredients for both schemes, in this section
we consider a matching to NNLO at the level of the cumulative cross
section that will allow us to estimate the systematic uncertainty
associated with the choice of the matching scheme. In this case the
resummed cross section is defined as in Eqs.~\eqref{eq:additive}
and~\eqref{eq:multiplicative1} with the obvious replacement of
${\rm N^3LO}$ by ${\rm NNLO}$.
The result of the comparison is reported in
Figure~\ref{fig:scheme_comparison}. We observe a very good agreement
between the two matching schemes, which is a sign of robustness of the
predictions shown below. The lower panel of
Figure~\ref{fig:scheme_comparison} shows the relative uncertainty
bands obtained within the two schemes, where each prediction is
divided by its own central value. The theory uncertainties have a very
similar pattern. Given that the difference between the two schemes is
always quite moderate with respect to the scale uncertainty, in the
following we decide to proceed with the multiplicative
prescription~\eqref{eq:multiplicative1} as our default. We find
analogous conclusions for DY production, and therefore we choose not
to report this further comparison here.

\subsection{Validation of the expansion for Drell-Yan pair production}
\label{sec:validationDY}

\begin{figure}[t]
  \centering
  \includegraphics[trim={0 -0.6cm 0 0},width=.45\linewidth]{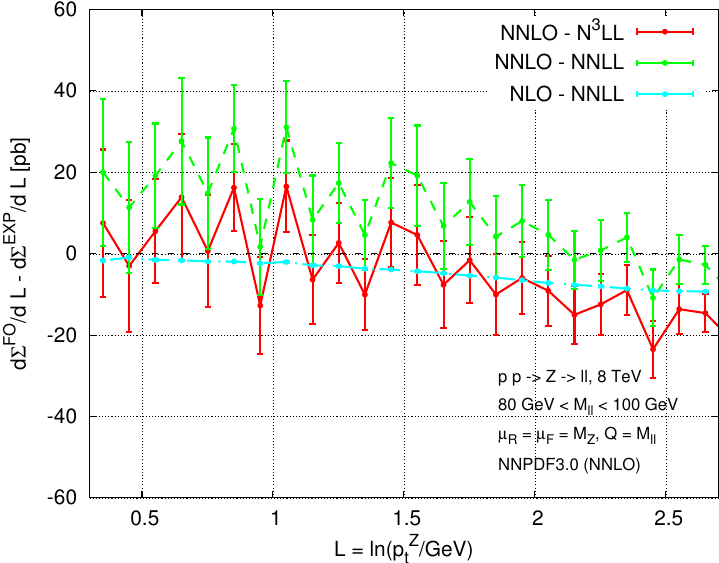}
  \qquad
  \includegraphics[width=.45\linewidth]{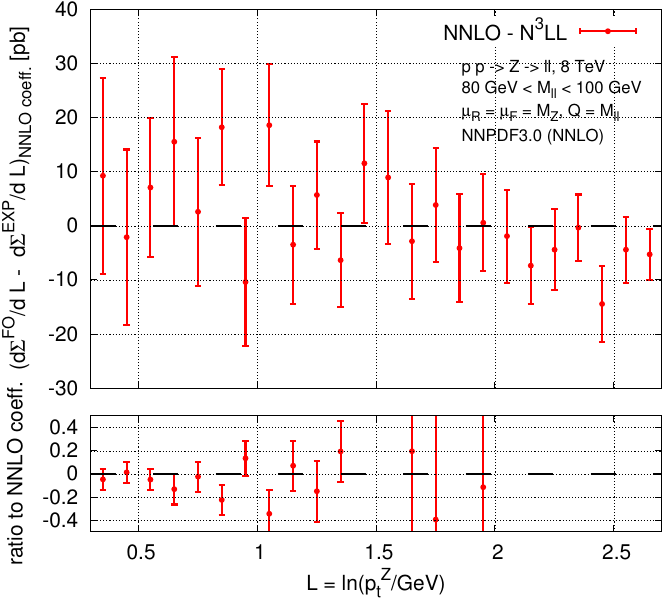} 
  \caption{Left: difference between the full NLO and NNLO $\ptz$
    distribution and the expansion of the NNLL and N$^3$LL
    resummation formulae~\eqref{eq:master-kt-space} to the 
    respective perturbative order. 
    Right: difference between the fixed-order NNLO coefficient,
    i.e.\ the $\order{\as^3}$ term alone, 
    and the corresponding coefficient obtained from the expansion of
    the N$^3$LL resummation.}
  \label{fig:res_v_FO_DY}
\end{figure}

\begin{figure}[htbp]
  \includegraphics[width=.48\linewidth]{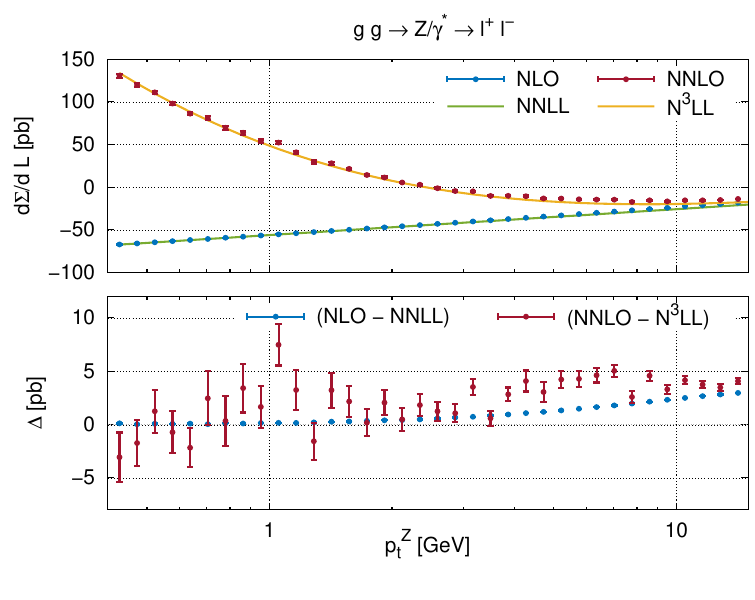}
  \hfill
  \includegraphics[width=.48\linewidth]{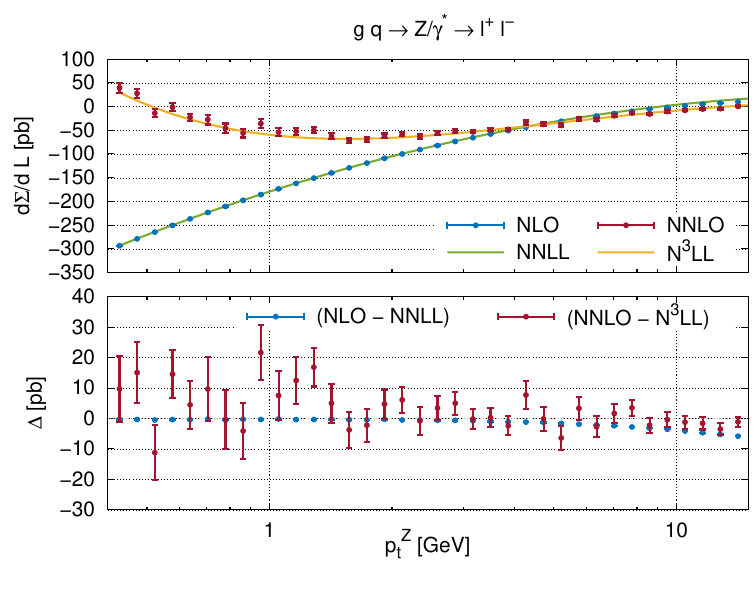}
  \\
  \includegraphics[width=.48\linewidth]{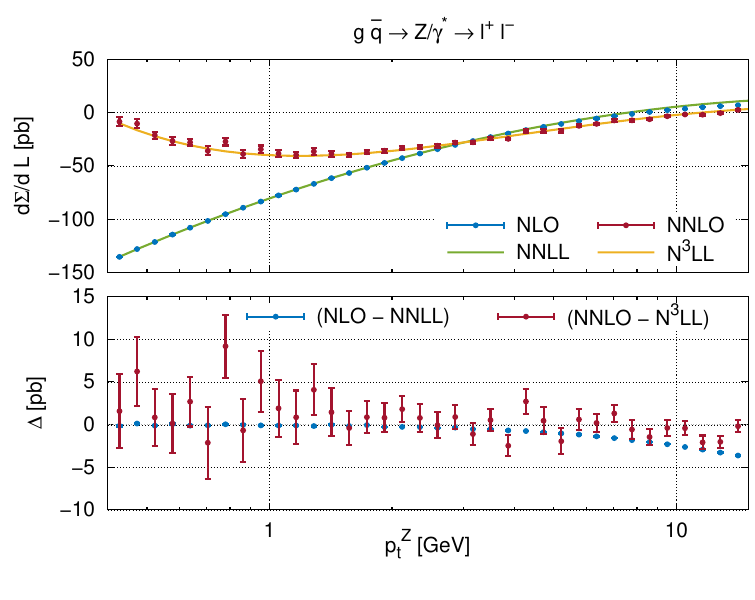}
  \hfill
  \includegraphics[width=.48\linewidth]{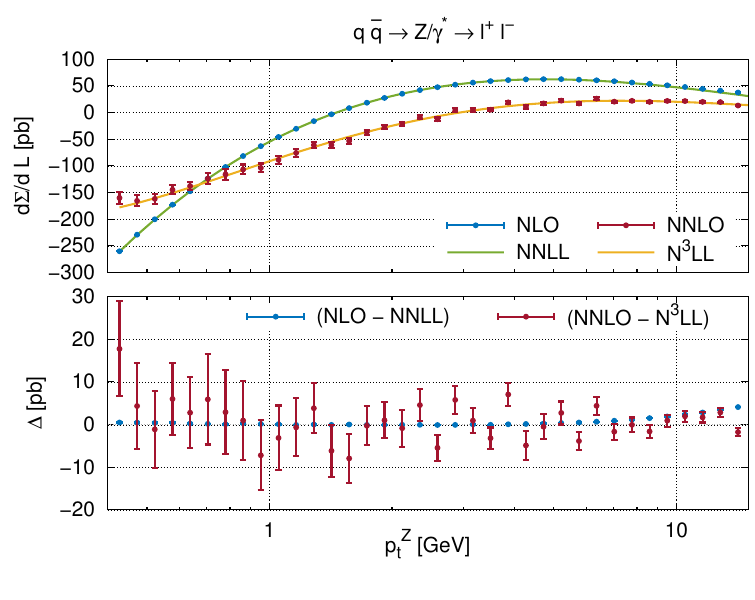}
  \\
  \includegraphics[width=.48\linewidth]{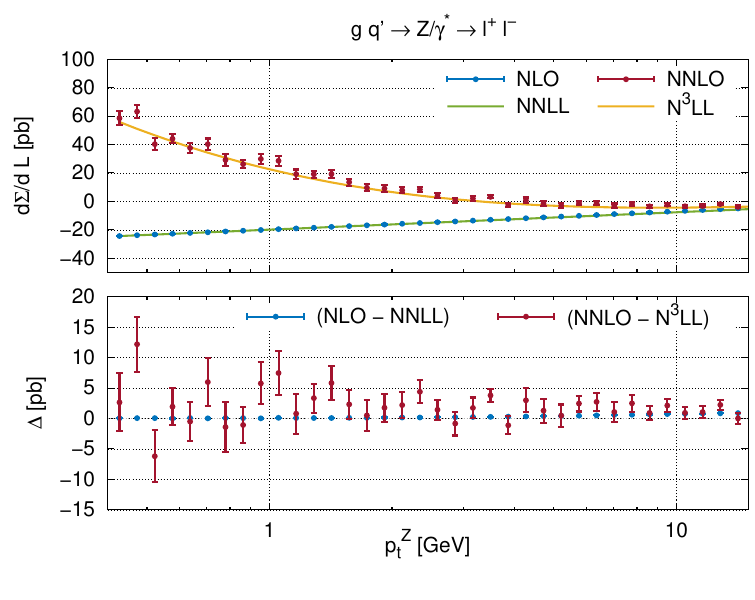}
  \hfill
  \includegraphics[width=.48\linewidth]{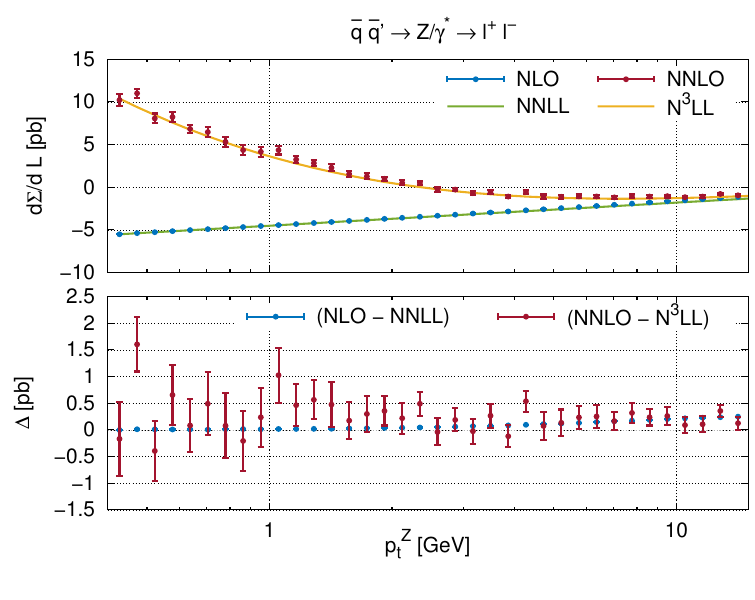}
  \caption{\label{fig:res_v_FO_DY_chan} Validation between the
    fixed-oder coefficients (at NLO and NNLO) and the corresponding
    expansion of the resummed prediction (at NNLL and N$^3$LL) for the
    individual partonic channels, with $L=\ln(\ptz/{\rm GeV})$.  Note
    that in contrast to Fig.~\ref{fig:res_v_FO_DY}, the curves
    labelled as ``NNLL'' only comprises term of $\order{\as^2}$ and
    does not include higher-order $\order{\as^3}$ terms.  }
\end{figure}

Similarly to the validation performed for inclusive Higgs production,
in this section we consider the difference between the NNLO
differential distribution and the corresponding expansion of the
N$^3$LL resummed calculation.  
In particular, we focus on the differential distribution
\begin{equation}
\label{eq:logdiffphs}
\frac{\rd \Sigma(\ptz)}{\rd\ln(\ptz/\GeV)} \, ,
\end{equation}
in order to highlight potential logarithmic differences in the
$\ptz\to0$ region. 

To perform the validation we consider $8~\TeV$ $pp$ collisions with
\texttt{NNPDF3.0} parton densities~\cite{Ball:2014uwa}, and we work
within an inclusive setup requiring
\begingroup
\setlength{\jot}{8pt}
\begin{align}
  80~\GeV & < \mll < 100~\GeV ,
\end{align}
\endgroup
  and setting the scales to $\mu_R=\mu_F=M_\mathrm{Z}$ with $x_Q=Q/\mll=1$.
  This inclusive setup is chosen as to avoid any potential complications
  due to the use of fiducial cuts, as well as dynamical scales,
  that act differently on the fixed-order and resummed
  calculations. Indeed, at variance with the case of the fixed-order
  calculation, in the resummation both fiducial cuts and dynamical
  scales are always defined at level of the Born (i.e.\ $\mathrm{Z}+0$\,jet) phase
  space, which differs from the definition used in the fixed-order
  calculation unless the extra QCD radiation is extremely soft or
  collinear to the beam. 
  As a consequence, employing fiducial cuts and/or dynamical scales
  may necessitate going
  to smaller values of $\ptz$ in order to see a convergence of the
  fixed-order to the expansion of the resummation.

 The results of the comparison are shown in Figure~\ref{fig:res_v_FO_DY}. The
  left panel displays the difference between the NLO distribution and
  the expansion of the NNLL resummation to second order (cyan dot-dashed line), and between
  the NNLO distribution and the expansion of the N$^3$LL resummation
  to third order (solid red line). In both cases one expects the differences to
  approach zero at small $\ptz$, which is well confirmed by the plot.
  In addition, we report on the difference between the NNLO distribution and the expansion of the NNLL resummation to third order given by the dashed green line.
  Due to missing double-logarithmic terms in the NNLL expansion, a non-vanishing slope is expected in the low-$\ptz$ region, which is suggested by the green curve within statical uncertainties.
  In order to single out the contribution of the NNLO correction, in
  the right panel of Figure~\ref{fig:res_v_FO_DY} we show the
  difference between the NNLO coefficient alone, and the corresponding
  coefficient in the expansion of the N$^3$LL resummation. As
  expected, such a difference asymptotically tends to zero for small
  $\ptz$ values.

  In addition to the validation of the full $\ptz$ spectrum shown in
  Fig.~\ref{fig:res_v_FO_DY}, we have further performed the analogous
  checks for the individual partonic channels which are summarised in
  Fig.~\ref{fig:res_v_FO_DY_chan}.  To this end, we have computed the
  fixed-order NNLO contribution to the $\ptz$ distribution down to
  $\ptz\sim0.5~\GeV$ with uncertainties at the $10\%$ level.  We can
  clearly observe that the fixed-order prediction is in excellent
  agreement with the prediction from the resummed calculation for all
  partonic configurations.  The respective bottom panels in each
  figure show the difference between the two predictions, which for
  all channels approach zero in the limit $\ptz\to0$.  This is an
  excellent cross-check of the two calculations, which proves the good
  numerical stability of the NNLO distributions down to the deep
  infrared regime.

\section{Results for Higgs production in HEFT}
\label{sec:Higgs}

In this section we present our predictions for the $\pth$ spectrum
both in inclusive $p p \to H$ production, and in the $p p \to H \to
\gamma \gamma$ channel with fiducial cuts. The computational setup
is the same for both analyses, and all results presented below are
obtained in the heavy-top-quark limit. 
We consider collisions at $13~\TeV$, and use parton densities from the
 {\tt PDF4LHC15\_nnlo\_mc}
set~\cite{Butterworth:2015oua,Dulat:2015mca,Ball:2014uwa,Harland-Lang:2014zoa,Carrazza:2015hva,Watt:2012tq}.
The value of the parameter $p$ appearing in the definition of the modified
logarithms
$\tilde L$ is chosen considering the scaling of the spectrum in the
hard region, so as to make the matching to the fixed order smooth
there. We set $p=4$ as our reference value, but nevertheless
have checked that a variation of $p$ by one unit does not induce any
significant differences.

We set the central renormalisation and factorisation scales as
$\mu_R=\mu_F=m_H/2$, with $m_H=125~\GeV$, while the resummation
scale is chosen to be $x_Q=Q/m_H=1/2$. We estimate the perturbative
uncertainty by performing a seven-scale variation of $\mu_R$, $\mu_F$
by a factor of two in either direction, while keeping
$1/2<\mu_R/\mu_F<2$ and $x_Q=1/2$; Moreover, for central $\mu_R$ and
$\mu_F$ scales, $x_Q$ is varied around its central value by a factor
of two. The quoted theoretical error is defined as the envelope of
all the above variations.
We discuss the results for inclusive production in
Section~\ref{sec:inclusive_h}, and then present the predictions for
the fiducial distributions in Section~\ref{sec:fiducial_h}.

\subsection{Matched predictions for inclusive Higgs}
\label{sec:inclusive_h}
\begin{figure}[tp]
  \centering
  \includegraphics[width=0.45\columnwidth]{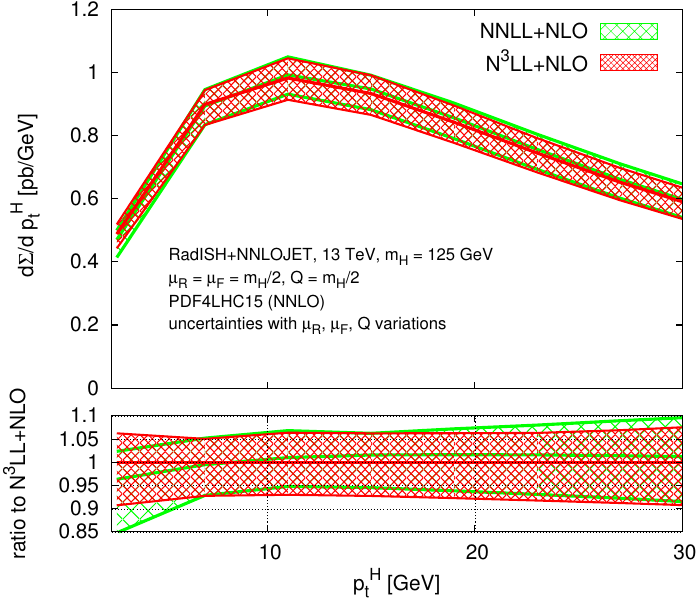} 
  \includegraphics[width=0.45\columnwidth]{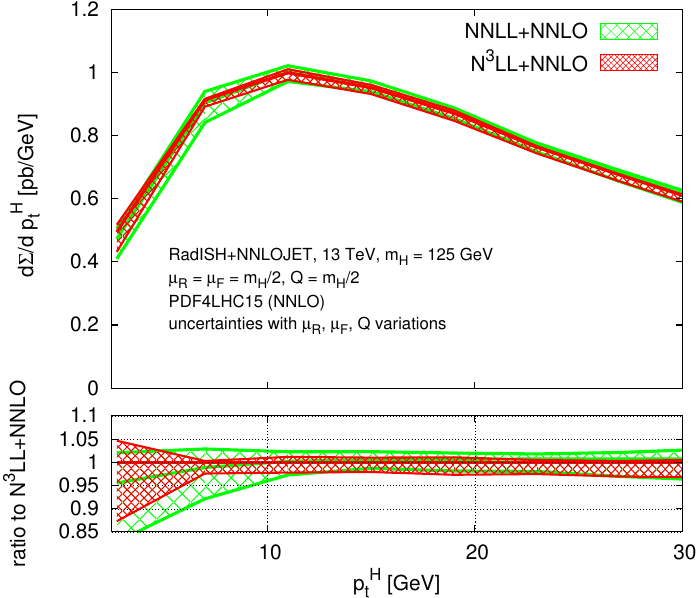} 
  \caption{Comparison between different combinations of fixed-order (NLO and NNLO) and resummation (NNLL and N$^3$LL) for the  transverse momentum distribution for Higgs boson production at $\sqrt{s} = 13~\TeV$. Left: NLO and Right: NNLO. The lower panel shows the ratio of predictions to that obtained with N$^3$LL resummation.}
  \label{fig:n3ll_v_nnll}
\end{figure}

We start by quantifying the size of the N$^3$LL effects compared to
NNLL resummation. In the left plot of Figure~\ref{fig:n3ll_v_nnll} we
compare the differential distributions at N$^3$LL+NLO and NNLL+NLO in
the small-$\pth$ region. The lower panel of the plot shows the ratio
of both predictions to the central line of the N$^3$LL+NLO band, which
corresponds to central scales in our setup. We observe that N$^3$LL
corrections are very moderate in size, with effects of order $2\%$ on
the central prediction in most of the displayed range, growing up to
at most $5\%$ only in the region of extremely low $\pth$. The central
N$^3$LL+NLO result is entirely contained in the NNLL+NLO uncertainty
band. On the other hand, the inclusion of the N$^3$LL corrections
reduces the perturbative uncertainty for $\pth \lesssim 5~\GeV$.

The right plot of Figure~\ref{fig:n3ll_v_nnll} shows the same
comparison for the matching to NNLO. The effect of the N$^3$LL
corrections is consistent with the previous order, with a
percent-level correction in most of the range, growing up to $5\%$ at
very small $\pth$. Similarly, the perturbative uncertainty is
significantly reduced below $10~\GeV$ with respect to the NNLL+NNLO
case.  It is important to stress that in the NNLL+NNLO matching the
fixed order and the expansion of the resummation differ by a divergent
term $\sim 1/\pth$ at small $\pth$. The fact that the divergence is
not visible in the distribution reported in the upper panel of
Figure~\ref{fig:n3ll_v_nnll} is entirely due to the nature of the
multiplicative scheme, which ensures that the distribution follows the
resummation scaling at small $\pth$, therefore damping the
divergence. A multiplicative matching of N$^3$LL resummation to NNLO
was already shown in Ref.~\cite{Bizon:2017rah}, where however no
significant reduction in the uncertainty band at small $\pth$ was
observed in that case. This feature was due to the limited statistics
of the fixed-order distributions used in that analysis at small
$\pth$, whose fluctuations dominated the uncertainty band at very
small transverse momentum. An additive matching of N$^3$LL to NNLO was
recently performed in Ref.~\cite{Chen:2018pzu}.

\begin{figure}[tp]
  \centering
  \includegraphics[width=0.45\columnwidth]{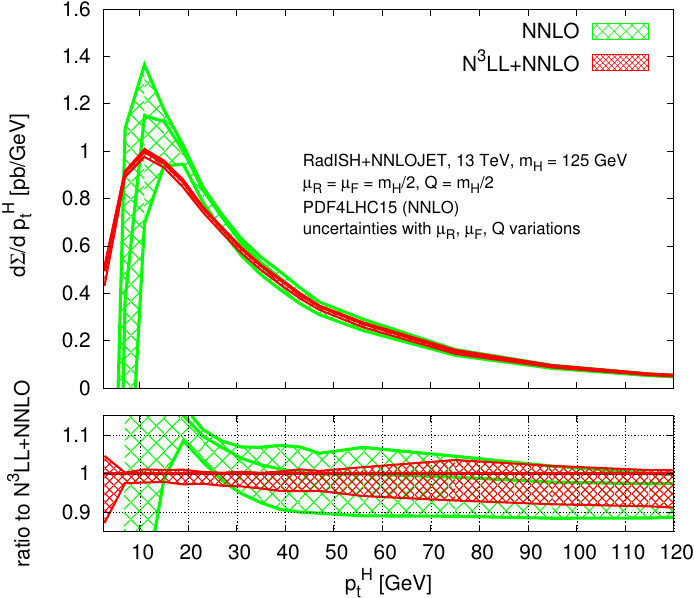} 
  \includegraphics[width=0.45\columnwidth]{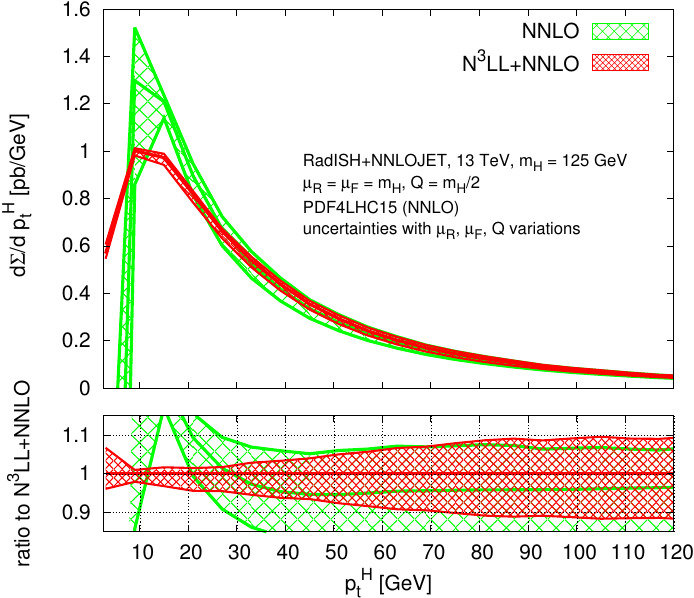} 
  \caption{Comparison of the  transverse momentum distribution for Higgs boson production at NNLO and N$^3$LL+NNLO for a central scale choice of $\mu_R = \mu_F =m_H /2$ (left) and $\mu_R = \mu_F = m_H$ (right).  In both cases, $Q=m_H/2$. The lower panel shows the ratio to the N$^3$LL+NNLO prediction.}
  \label{fig:n3ll_nnlo_v_nnlo}
\end{figure}

Next, we consider the comparison between the matched prediction and the
fixed-order one. Figure~\ref{fig:n3ll_nnlo_v_nnlo} shows this
comparison for two different central scales. The left plot is obtained
with central $\mu_F=\mu_R= m_H/2$, while the right plot is obtained
with $\mu_F=\mu_R= m_H$. The rest of the setup is kept as described
above. We observe that at $\mu_F=\mu_R= m_H/2$ the uncertainty band is
affected by cancellations in the scale variation, which accidentally
lead to a small perturbative uncertainty. Choosing $m_H$ as a central
scale (right plot of Figure~\ref{fig:n3ll_nnlo_v_nnlo}) leads to a
broader uncertainty band resulting in a more robust estimate of the
perturbative error. This is particularly the case for predictions
above $50~\GeV$, where resummation effects are progressively less
important. We notice indeed that in both cases the effect of
resummation starts to be increasingly relevant for
$\pth\lesssim 40~\GeV$.

In the following we choose $m_H/2$ as a central scale. Nevertheless,
we stress that a comparison to data (not performed here for Higgs
boson production) will require a study of different central-scale
choices.

\begin{figure}[tp]
  \centering
  \includegraphics[width=0.45\columnwidth]{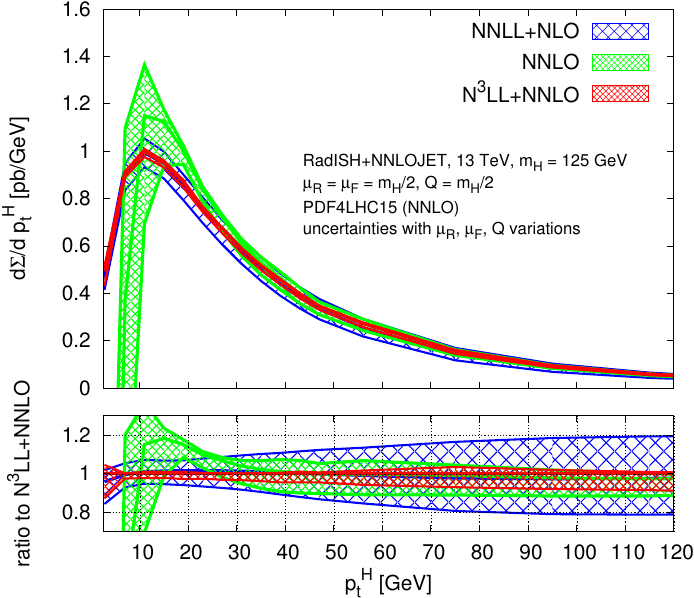} 
  \caption{Comparison of the  transverse momentum distribution for Higgs boson production between N$^3$LL+NNLO, NNLL+NLO, and NNLO at central scale choice of $\mu_R = \mu_F =m_H /2$. The lower panel shows the ratio to the N$^3$LL+NNLO prediction.}
  \label{fig:n3ll_nnlo_v_nnlo_final}
\end{figure}

To conclude, Figure~\ref{fig:n3ll_nnlo_v_nnlo_final} reports the
comparison between our best prediction (N$^3$LL+NNLO), the
NNLL+NLO, and the NNLO distributions. The plot shows a very good
convergence of the predictions at different perturbative orders, with
a significant reduction of the scale uncertainty in the whole
kinematic range considered here.

\subsection{Matched predictions for fiducial \texorpdfstring{$H\to \gamma\gamma$}{H -> gamma gamma}}
\label{sec:fiducial_h}
Experimental measurements are performed within a fiducial phase-space
volume, defined in order to comply with the detector geometry and to
enhance signal sensitivity. On the theoretical side it is therefore
highly desirable to provide predictions that exactly match the experimental
setup. The availability of matched predictions that are fully
differential in the Born phase space also allows for a direct
comparison to data without relying on Monte Carlo modeling of
acceptances. In this section we consider the process $pp\to H\to
\gamma\gamma$ and, in particular, we focus on the transverse momentum
of the $\gamma\gamma$ system in the presence of fiducial cuts.

The fiducial volume is defined by the set of cuts detailed below~\cite{Aaboud:2018xdt}
\begingroup
\setlength{\jot}{8pt}
\begin{align}
\label{eq:Higgs_fiducial}
\min(\ptgo,\ptgt)> 31.25~\GeV,\qquad \max(\ptgo,\ptgt)> 43.75~\GeV,
\notag\\
0<|\eta^{\gamma_{1,2}}|<1.37 ~~{\rm or}~~ 1.52<|\eta^{\gamma_{1,2}}|<2.37,\qquad |Y_{\gamma\gamma}|<2.37
\,,
\end{align}
\endgroup
where $\ptgo$, $\ptgt$ are the transverse momenta of the two photons,
$\etag$ are their pseudo-rapidities in the hadronic centre-of-mass
frame, and $Y_{\gamma\gamma}$ is the photon-pair rapidity. In the
definition of the fiducial volume we do not include the
photon-isolation requirement, since this would introduce additional
logarithmic corrections of non-global nature in the problem, spoiling
the formal N$^3$LL+NNLO accuracy of the differential
distributions.\footnote{However, we point out that photon-isolation
  criteria in this case are not aggressive, and therefore they could
  be safely included at fixed order.}  We consider on-shell Higgs
boson production followed by a decay into two photons under the
narrow-width approximation with a branching ratio of
$2.35\times 10^{-3}$.

\begin{figure}[tp]
  \centering
  \includegraphics[width=0.45\columnwidth]{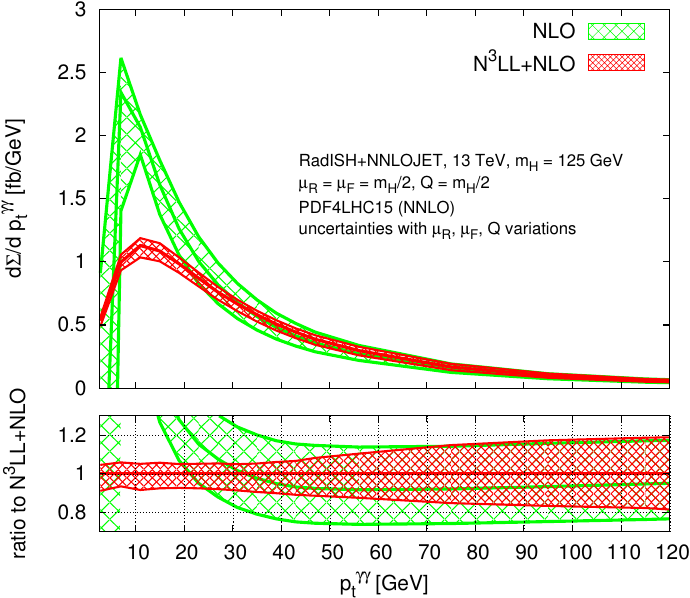} 
  \includegraphics[width=0.45\columnwidth]{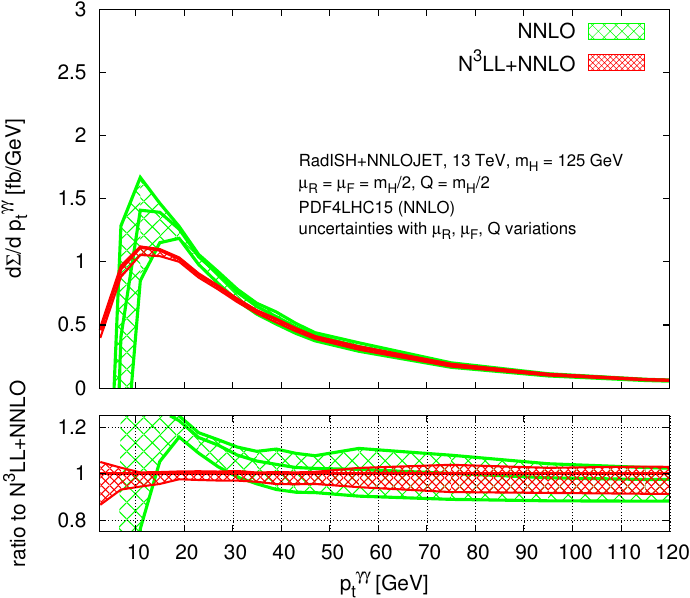} 
  \caption{Comparison of the transverse momentum distribution for Higgs boson production at $\sqrt{s} = 13~\TeV$ in the fiducial volume defined by Eq.~\eqref{eq:Higgs_fiducial} at N$^3$LL+NLO and NLO (left) and N$^3$LL+NNLO and NNLO (right). The lower panel shows the ratio to the N${}^3$LL+NNLO prediction.}
  \label{fig:n3ll_nnlo_v_nnlo_gamgam}
\end{figure}

In Figure~\ref{fig:n3ll_nnlo_v_nnlo_gamgam} we show the comparison of
the matched and the fixed-order predictions for the transverse momentum
of the photon pair in the fiducial volume, at different perturbative
accuracies: N$^3$LL+NLO vs.\ NLO in the left panel, and N$^3$LL+NNLO
vs.\ NNLO in the right one.

By comparing the two panels of Figure~\ref{fig:n3ll_nnlo_v_nnlo_gamgam}
we notice a substantial reduction in the theoretical uncertainty in the
medium-high-$\ptgg$ region, driven by the increase in perturbative accuracy
of the fixed-order computation; at very low $\ptgg$, the
prediction is dominated by resummation, which is common
to both panels. The pattern observed in the right panel is very similar to
what we obtained in the inclusive case in the left panel of
Figure~\ref{fig:n3ll_nnlo_v_nnlo}. We stress again that the particularly
small uncertainty of the matched prediction is to a certain extent due to
the choice of central scales we adopt, namely $\mu_R=\mu_F=m_H/2$, which
suffers from large accidental cancellations.

\FloatBarrier

\section{Results for Drell-Yan production}
\label{sec:DY}
We now turn to the study of Drell-Yan pair production at the LHC. In
this section we present the results for the differential distributions
of the transverse momentum of the DY pair, as well as for the angular
observable $\phs$.

We consider $8~\TeV$ proton-proton collisions, and compare the resulting
calculation for the differential spectra with ATLAS data from
Ref.~\cite{Aad:2015auj}. The fiducial phase-space volume is defined as
follows:
\begingroup
\setlength{\jot}{8pt}
\begin{align}
  \label{eq:DY_fiducial}
  p_{t}^{\ell^\pm}   & > 20~\GeV, & 
  |\eta^{\ell^\pm}|  & < 2.4 , &
  |\yll| &< 2.4, &
  46~\GeV & < \mll < 150~\GeV ,
\end{align}
\endgroup where $p_{t}^{\ell^\pm}$ are the transverse momenta of the two
leptons, $\eta^{\ell^\pm}$ are their pseudo-rapidities, while $\yll$ and
$\mll$ are the rapidity and invariant mass of the di-lepton
system, respectively. All rapidities and pseudo-rapidities are evaluated in the
hadronic centre-of-mass frame.

For our results, we use parton densities as obtained from the
\texttt{NNPDF3.0} set. The reference value we set for the parameter
$p$ appearing in the modified logarithms is $p=4$, but we have checked
that a variation of $p$ by one unit does not induce any significant
differences.

We set the central scales as $\mu_R=\mu_F=M_T=\sqrt{\mll^2+(\ptz)^2}$,
while the central resummation scale is chosen to be
$x_Q=Q/\mll=1/2$. The theoretical uncertainty is estimated through the
same set of variations as for Higgs boson production.

The results for $\ptz$ and $\phs$ are shown in the following two
subsections. All plots have the same pattern: the main panels display
the comparison of normalised differential distributions at NNLO
(green), NNLL+NLO (blue), and N$^3$LL+NNLO (red), respectively,
overlaid on ATLAS data points (black). Correspondingly, the lower
insets of each panel show the ratio of the theoretical curves to data,
with the same colour code as in the main panels.

\subsection{Matched predictions for fiducial \texorpdfstring{$\ptz$}{ptz} distributions}
\label{sec:ptz-results}

\begin{figure}[tp]
  \centering
  \includegraphics[width=0.45\columnwidth]{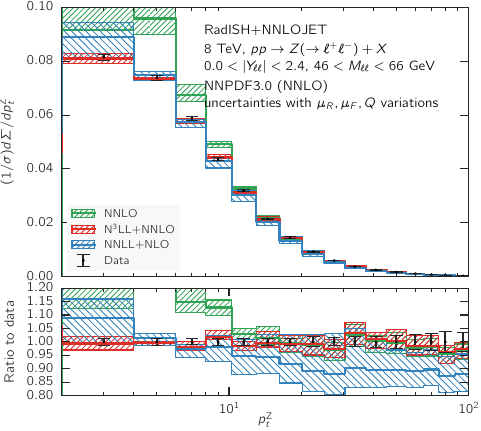} 
  \includegraphics[width=0.45\columnwidth]{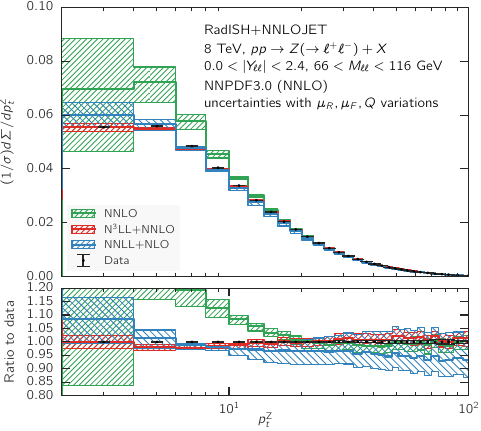}
  \includegraphics[width=0.45\columnwidth]{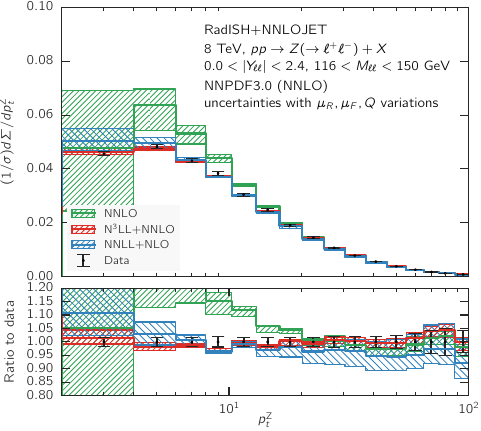} 
  \caption{Comparison of the normalised  transverse momentum distribution for Drell-Yan pair production at NNLO (green), NNLL+NLO (blue) and N$^3$LL+NNLO (red) at $\sqrt{s} = 8~\TeV$ integrated over the full lepton-pair rapidity range ($0 < |Y _{\ell\ell} | < 2.4$), in three different lepton-pair invariant-mass windows. For reference, the ATLAS data is also shown, and the lower panel shows the ratio of each prediction to data.}
  \label{fig:dy_ptz_minv_windows_fullY}
\end{figure}

%%%%%%%%%%%%%%%%%%%%%%%%%
In Figure~\ref{fig:dy_ptz_minv_windows_fullY} we display the normalised $\ptz$ 
distributions in which, in addition to the fiducial cuts
reported above, we consider three different lepton-pair invariant-mass windows:
\begin{align}
  \label{eq:dy_minv_windows}
  &{\rm low~ invariant~ mass:}     &\hspace{-1cm} ~46~\GeV < \mll < ~66~\GeV, \notag\\
  &{\rm medium~ invariant~ mass:}  &\hspace{-1cm} ~66~\GeV < \mll < 116~\GeV, \notag\\
  &{\rm high~ invariant~ mass:}    &\hspace{-1cm} 116~\GeV < \mll < 150~\GeV.
\end{align}

A comparison of the most accurate matched prediction with the
fixed-order one shows that the N$^3$LL+NNLO prediction starts differing
significantly from the NNLO for $\ptz \lesssim 15~\GeV$, while for
$\ptz > 20~\GeV$ the NNLO is sufficient to provide a reliable
description. Comparing matched predictions with different formal
accuracy, we note that the N$^3$LL+NNLO curve has a significantly reduced
theoretical systematics with respect to that for NNLL+NLO, in the whole
$\ptz$ range and for all considered invariant-mass windows. The
perturbative error is reduced by more than a factor of two at very low
$\ptz$, where the prediction is dominated by resummation, and the
leftover uncertainty in that region is as small as $3$--$5\%$, and almost
comparable with the excellent experimental precision.  The shape of
the $\ptz$ distributions is also significantly distorted by the
inclusion of higher orders: the spectrum is harder than the NNLL+NLO
result for $\ptz \gtrsim 10~\GeV$, and the peak is lower, with the
N$^3$LL+NNLO curves in much better agreement with data with respect to
NNLL+NLO in the whole kinematic range. Among the three considered
windows, the most accurately described at N$^3$LL+NNLO are the ones at
intermediate and high invariant mass; the accuracy very slightly
degrades for smaller invariant masses, however the theory uncertainty
never gets larger than $5$--$7\%$ over the whole displayed $\ptz$ range.

\begin{figure}[tp]
  \centering
  \includegraphics[width=0.45\columnwidth]{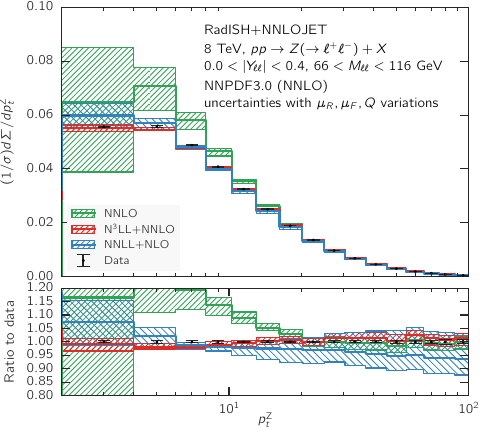}
  \includegraphics[width=0.45\columnwidth]{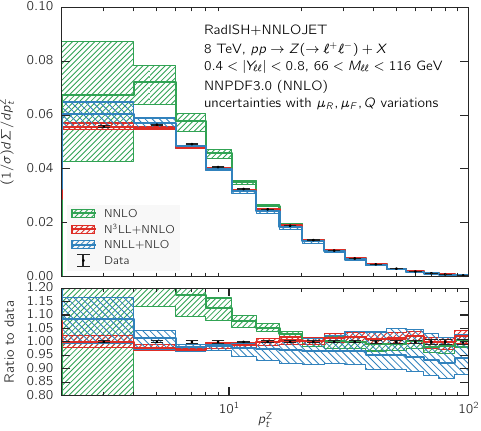} 
  \includegraphics[width=0.45\columnwidth]{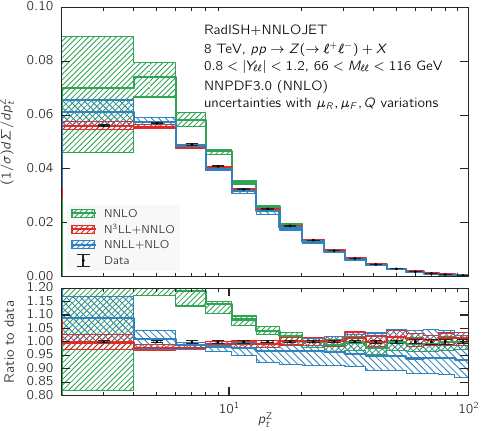} 
  \includegraphics[width=0.45\columnwidth]{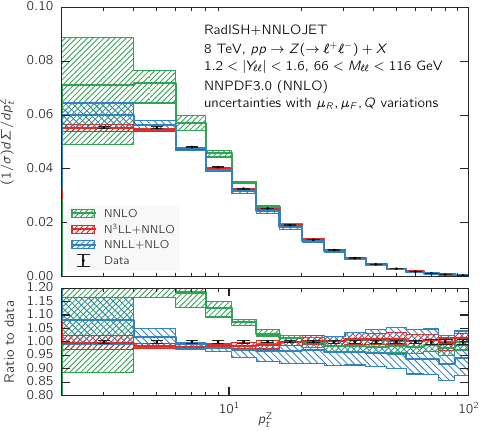}   
  \includegraphics[width=0.45\columnwidth]{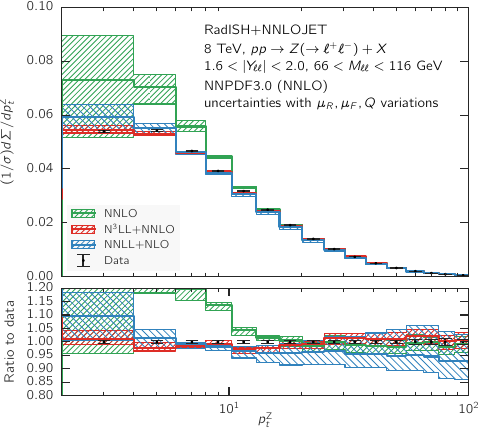} 
  \includegraphics[width=0.45\columnwidth]{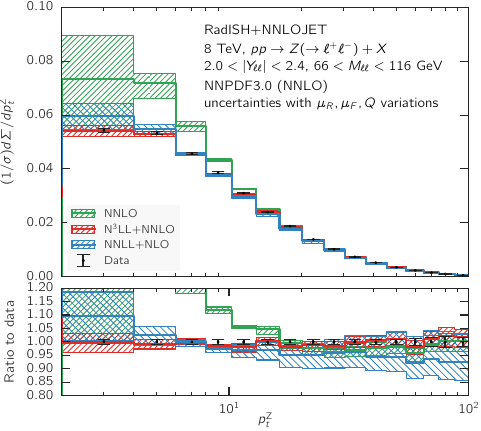} 
  \caption{Comparison of the normalised  transverse momentum distribution for Drell-Yan pair production at NNLO (green), NNLL+NLO (blue) and N$^3$LL+NNLO (red) at $\sqrt{s} = 8~\TeV$ in the central lepton-pair invariant-mass window ($66~\GeV< \mll < 116~\GeV$) for six different lepton-pair rapidity slices. For reference, the ATLAS data is also shown, and the lower panel shows the ratio of each prediction to data.}
  \label{fig:dy_ptz_cent_minv_Yslices}
\end{figure}

%%%%%%%%%%%%%%%%%%%%%%%%%
In Figure~\ref{fig:dy_ptz_cent_minv_Yslices} we focus our analysis on
the central lepton-pair invariant-mass window defined in
Eq.~\eqref{eq:dy_minv_windows} and show predictions for the normalised
$\ptz$ distribution in six different lepton-pair rapidity slices:
\begin{align}
  \label{eq:dy_rapidity_slices}
  \text{(a)}~&      0.0 < \lvert \yll \rvert < 0.4, &
  \text{(b)}~&      0.4 < \lvert \yll \rvert < 0.8, &
  \text{(c)}~&      0.8 < \lvert \yll \rvert < 1.2, \notag\\
  \text{(d)}~&      1.2 < \lvert \yll \rvert < 1.6, &
  \text{(e)}~&      1.6 < \lvert \yll \rvert < 2.0, &
  \text{(f)}~&      2.0 < \lvert \yll \rvert < 2.4.
\end{align}

The comments relevant to Figure~\ref{fig:dy_ptz_minv_windows_fullY} by
far and large apply in this case as well, with our best prediction at
N$^3$LL+NNLO affected by an uncertainty that is of order $3$--$5\%$ in
the whole $\ptz$ range, regardless of the considered rapidity
slice. It is moreover in very good agreement with the experimental
data, hence significantly improving on both the NNLL+NLO, in the whole
$\ptz$ range, and the pure NNLO, in the $\ptz\lesssim 20~\GeV$ region.

%\newpage
\subsection{Matched predictions for fiducial \texorpdfstring{$\phs$}{phistar} distributions}
\label{sec:phs-results}

\begin{figure}[tp]
  \centering
  \includegraphics[width=0.45\columnwidth]{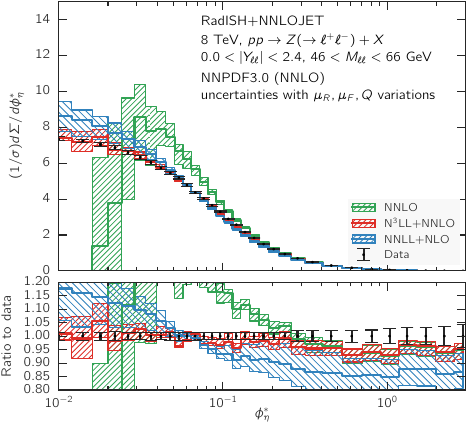}
  \includegraphics[width=0.45\columnwidth]{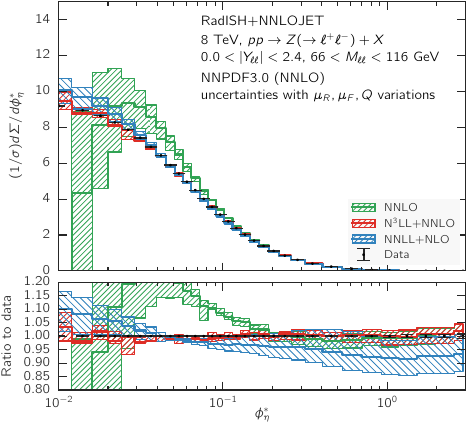}
  \includegraphics[width=0.45\columnwidth]{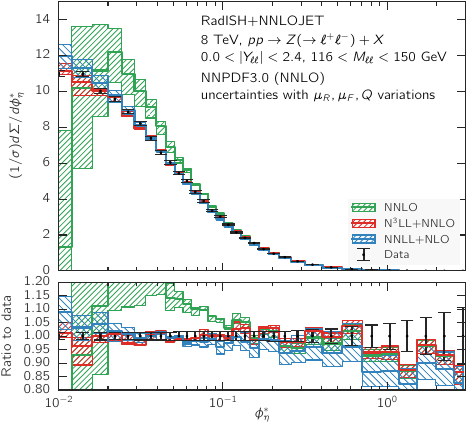}
  \caption{Comparison of the normalised $\phs$ distribution for
    Drell-Yan pair production at NNLO (green), NNLL+NLO (blue) and
    N$^3$LL+NNLO (red) at $\sqrt{s} = 8~\TeV$ integrated over the full
    lepton-pair rapidity range ($0 < |Y _{\ell\ell} | < 2.4$), in
    three different lepton-pair invariant-mass windows. For reference,
    the ATLAS data is also shown, and the lower panel shows the ratio
    of each prediction to data.}
  \label{fig:dy_phs_minv_windows_fullY}
\end{figure}

Figure~\ref{fig:dy_phs_minv_windows_fullY} shows the $\phs$
distribution for three different lepton-pair invariant-mass windows as defined in Eq.~\eqref{eq:dy_minv_windows}.

The pattern of comparisons among theoretical predictions is
qualitatively similar to what discussed for the $\ptz$
distribution. Resummation effects at N$^3$LL+NNLO start being
important with respect to the pure NNLO in the region
$\phs \lesssim 0.2$; the shape of the N$^3$LL+NNLO distribution is
significantly distorted with respect to the NNLL+NLO one in a
similar fashion as for the $\ptz$ case, and the uncertainty band is
reduced by a factor of two or more over the whole range and for all
invariant-mass windows, down to the level of $3$--$5\%$ (except at low
invariant mass, where the uncertainty is $5$--$7\%$).

At variance with the $\ptz$ case, however, for $\phs$ we note that the
N$^3$LL+NNLO prediction describes data appropriately only in the
central- and high- invariant-mass windows. In the low-invariant-mass
one, the prediction undershoots data in the medium-hard region, by up
to $5$--$7\%$. This tension was already observed in the fixed-order NNLO
comparison~\cite{Gehrmann-DeRidder:2016jns}. However, given the large
statistical uncertainty of the data in this invariant-mass range, the
theory still provides a reasonable description of the measurement, and
the N$^3$LL+NNLO prediction is in much better agreement with data than
the NNLL+NLO in the whole range of $\phs$, especially at low $\phs$.

\begin{figure}[tp]
  \centering
  \includegraphics[width=0.45\columnwidth]{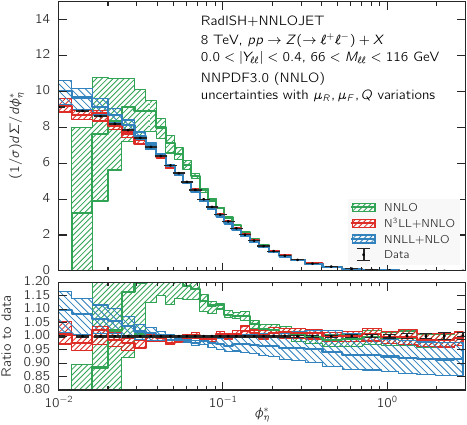}
  \includegraphics[width=0.45\columnwidth]{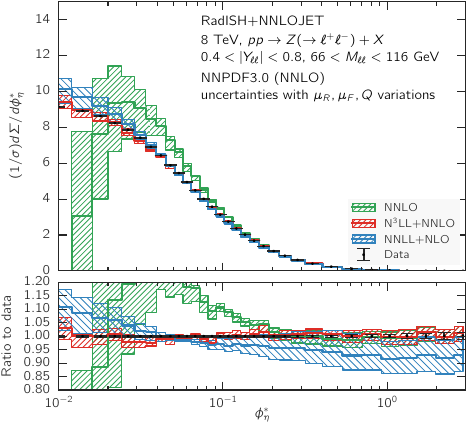}
  \includegraphics[width=0.45\columnwidth]{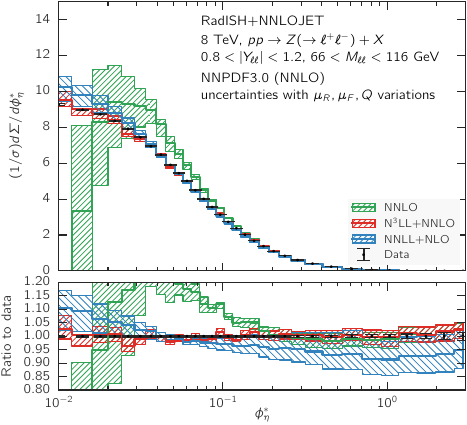}
  \includegraphics[width=0.45\columnwidth]{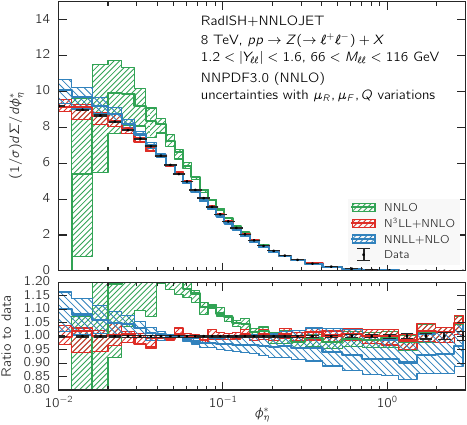}
  \includegraphics[width=0.45\columnwidth]{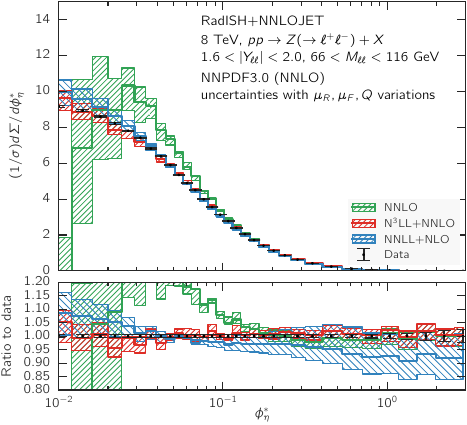}
  \includegraphics[width=0.45\columnwidth]{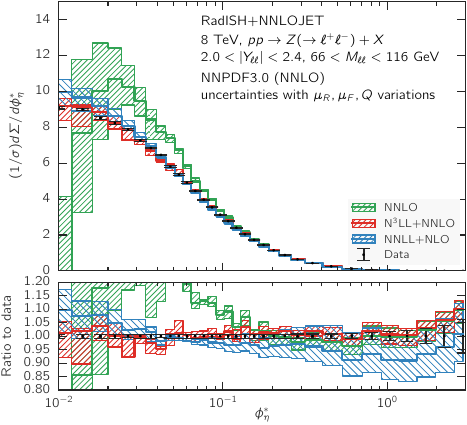}
  \caption{Comparison of the normalised  $\phs$
    distribution for Drell-Yan pair production  at NNLO (green),
    NNLL+NLO (blue) and N$^3$LL+NNLO (red) at $\sqrt{s} = 8~\TeV$ in
    the central lepton-pair invariant-mass window ($66~\GeV < \mll < 116~\GeV$) for three different lepton-pair rapidity slices. For reference, the ATLAS data is also shown, and the lower panel shows the ratio of each prediction to data.}
  \label{fig:dy_phs_cent_minv_Yslices}
\end{figure}

\begin{figure}[tp]
  \centering
  \includegraphics[width=0.45\columnwidth]{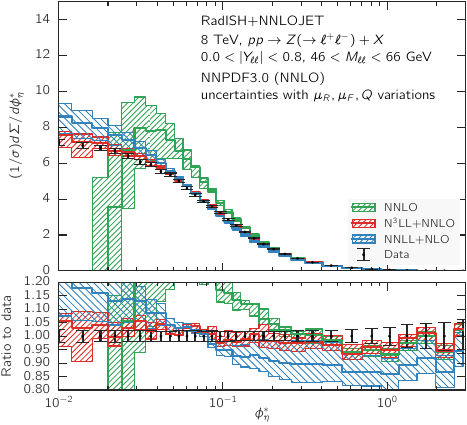}
  \includegraphics[width=0.45\columnwidth]{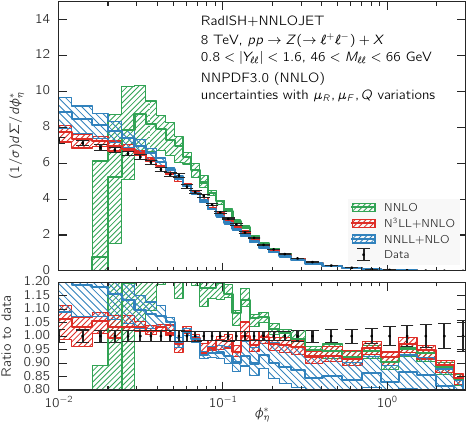}
  \includegraphics[width=0.45\columnwidth]{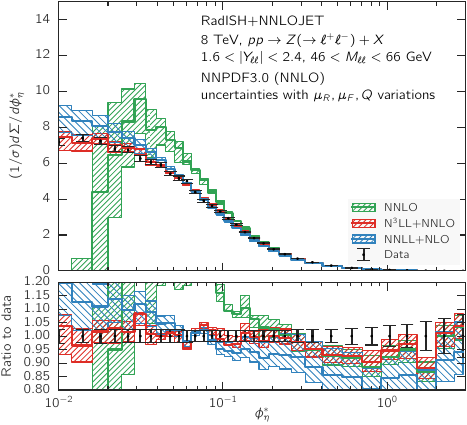}
  \caption{Comparison of the normalised  $\phs$ distribution for
    Drell-Yan pair production  at NNLO (green), NNLL+NLO (blue) and N$^3$LL+NNLO (red) at $\sqrt{s} = 8~\TeV$ in the low lepton-pair invariant-mass window ($46~\GeV < \mll < 66~\GeV$) for three different lepton-pair rapidity slices. For reference, the ATLAS data is also shown, and the lower panel shows the ratio of each prediction to data.}
  \label{fig:dy_phs_low_minv_Yslices}
\end{figure}

\begin{figure}[tp]
  \centering
  \includegraphics[width=0.45\columnwidth]{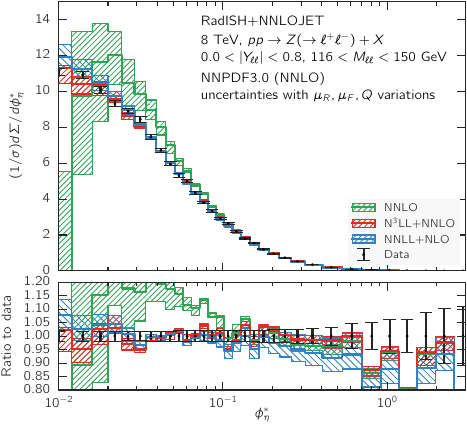}
  \includegraphics[width=0.45\columnwidth]{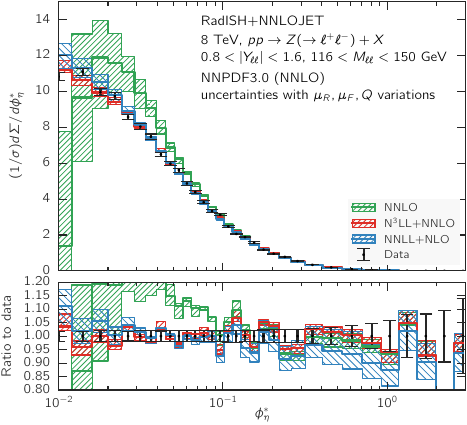}
  \includegraphics[width=0.45\columnwidth]{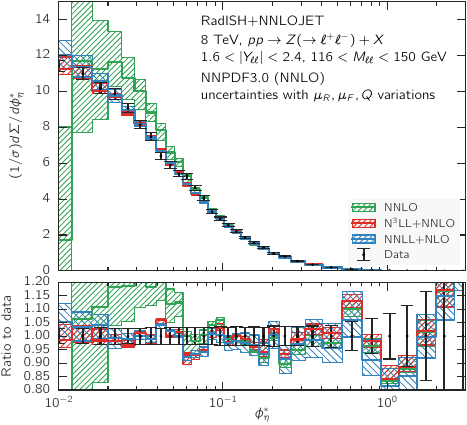}
  \caption{Comparison of the normalised  $\phs$ distribution for
    Drell-Yan pair production  at NNLO (green), NNLL+NLO (blue) and N$^3$LL+NNLO (red) at $\sqrt{s} = 8~\TeV$ in the high lepton-pair invariant-mass window ($116~\GeV < \mll < 150~\GeV$) for three different lepton-pair rapidity slices. For reference, the ATLAS data is also shown, and the lower panel shows the ratio of each prediction to data.}
  \label{fig:dy_phs_high_minv_Yslices}
\end{figure}

In Figure~\ref{fig:dy_phs_cent_minv_Yslices} we show the results for the $\phs$ distributions in the central invariant-mass window, see Eq.~\eqref{eq:dy_minv_windows}, split into the six lepton-pair rapidity slices described in Eq.~\eqref{eq:dy_rapidity_slices}. Moreover, given the availability of experimental measurements, in Figures~\ref{fig:dy_phs_low_minv_Yslices} and \ref{fig:dy_phs_high_minv_Yslices} we also provide predictions sliced in $\yll$ for the low- and high- di-lepton invariant-mass windows, respectively. The three rapidity slices we focus on correspond to regions (a+b), (c+d), and (e+f) of Eq.~\eqref{eq:dy_rapidity_slices}.

The prediction subdivided in rapidity slices largely shares the same
features as that integrated over rapidity, which has been detailed
in Figure~\ref{fig:dy_phs_minv_windows_fullY}. In the central
invariant-mass window, data is accurately reproduced by the
N$^3$LL+NNLO prediction, regardless of the considered rapidity slice,
with a theoretical systematics in the $5\%$ range or smaller. 
The quality of the description slightly degrades at low invariant
mass, and to a lesser extent also at high invariant mass, mainly in
the hard region, with a pattern similar to that displayed by the
rapidity-integrated spectrum.  Overall, the uncertainty associated
with the N$^3$LL+NNLO is of order of $5\%$ or better, with a
significant improvement both in the shape and in the systematics with
respect to NNLL+NLO.


\FloatBarrier

%\newpage
\section{Conclusions}
\label{sec:conclusions}

In this work we have presented precise predictions for differential distributions in Higgs boson and Drell-Yan pair production at the LHC at N$^3$LL+NNLO.

The resummation is performed in momentum space and is fully exclusive in the Born phase space.
For the matching to NNLO we adopted a multiplicative scheme, which
allows for the inclusion of the N$^3$LO constant terms to the
cumulative cross section. These are currently unknown analytically,
but can be included numerically once the total N$^3$LO cross section
has been obtained.
The uncertainty associated with the choice of the
matching scheme was estimated at NLO accuracy, for which an additive
matching with the same ingredients can be also performed. At this
order the predictions obtained with the two prescriptions are in very
good agreement with each other, and the matching-scheme uncertainty is
under control within the perturbative error.

For Higgs boson production in gluon fusion, we have considered the transverse-momentum spectrum both at the inclusive level and in the $H\rightarrow \gamma \gamma$ channel within ATLAS fiducial cuts.
In both cases, we observe that the resummation reduces the theoretical
uncertainties and stabilises the fixed-order result below
$\pth \sim 40~\GeV$.
The effects of the N$^3$LL corrections with respect to NNLL+NNLO distributions
are moderate in size, with a percent-level correction in most of the
range, growing up to $5\%$ at very small $\pth$. However, the
perturbative uncertainty is reduced significantly below $10~\GeV$
with respect to the NNLL+NNLO case.

For Drell-Yan pair production, we have presented resummed predictions within ATLAS fiducial cuts~\cite{Aad:2015auj} both for the normalised $\ptz$ distributions and for the normalised $\phs$ distributions, and we have compared them to experimental data.
In the case of transverse-momentum distributions, the difference
between the fixed-order and the N$^3$LL+NNLO result becomes
significant for $\ptz < 10$--$15~\GeV$, while for $\ptz > 20~\GeV$ the
NNLO prediction is sufficient to provide a reliable description of the
experimental data. Comparing matched results with different formal
accuracy, we note that the N$^3$LL+NNLO prediction has a significantly reduced
theoretical uncertainty with respect to that for NNLL+NLO, in the whole
$\ptz$ range and for all invariant-mass windows considered
in our study.

For the $\phs$ distribution, resummation effects start being important
with respect to pure NNLO in the region $\phs \lesssim 0.2$. At
N$^3$LL+NNLO the shape of the distribution is significantly distorted
with respect to that for NNLL+NLO (the spectrum is hardened in
the tail, and the height of the peak is lowered), and the uncertainty
band is reduced by a factor of two or more over the whole range of
$\phs$ and for most invariant-mass windows, down to the level of
$3$--$5\%$. An exception is at low invariant mass, where
the uncertainty remains in the $5$--$7\%$ range. Unlike the $\ptz$
case, for $\phs$ we note that the N$^3$LL+NNLO prediction describes
data appropriately only in the central- and high-invariant-mass
windows, while at low invariant mass the prediction undershoots the data
in the medium-hard region. The difference between the central values
of the data and theory here can be of the order of $10\%$, however no
significant tension with the data is observed, due to the sizeable
statistical uncertainty in the measurement. The agreement in these
invariant-mass bins is much improved by the inclusion of the
N$^3$LL+NNLO corrections with respect to the NNLL+NLO distribution.

Our results are an important step in the LHC precision programme, where
accurate predictions have become necessary for an appropriate
interpretation and exploitation of data. 
In order to improve on the predictions presented here, several effects
must be considered.

For Higgs boson production via gluon fusion, the impact of other heavy quarks, notably the bottom quark,
becomes relevant at this level of
accuracy and therefore must be taken into account. Recent studies show
that the effect of the top-bottom interference at
NNLL+NLO~\cite{Lindert:2017pky,Caola:2018zye} could lead to
distortions of the transverse-momentum spectrum that are as large as
$\sim 5\%$ with respect to the HEFT approximation, and the theory
uncertainties associated with this contribution are
of ${\cal O}(20\%)$. These effects are therefore of the same order as
the perturbative uncertainties presented here, and must be included for
a consistent prediction of the spectrum with $5$--$10\%$ perturbative
accuracy in the region $\pth \lesssim m_H$.

In the DY case, the situation is more involved given the smaller
perturbative uncertainty. At this level of precision, it is necessary
to supplement the predictions obtained in this work at small $\ptz$
and $\phs$ with QED corrections and with an estimate of various
sources of non-perturbative effects that could be as large
as a few $\%$ in this region. Similarly, the inclusion of quark masses
may have a few-percent effect on the
spectrum~\cite{Pietrulewicz:2017gxc,Bagnaschi:2018dnh}, and more
precise studies are necessary in order to assess their impact
precisely. Recent analyses~\cite{Bagnaschi:2018dnh} suggest that the
inclusion of these effects may have a non-negligible impact on
observables of current phenomenological interest, such as the
determination of the $W$-boson mass~\cite{Aaboud:2017svj}. Given that
the size of these effects is of the order of the perturbative
uncertainty of the N$^3$LL+NNLO prediction, a careful assessment will
be necessary to improve further on the results presented in this work.

\acknowledgments 
XC and TG thank the University of Zurich S3IT for providing
 the computational resources for this project. XC, TG and AH acknowledge the computing resources provided by the Swiss National Supercomputing Centre (CSCS) under the project ID p501b and UZH10.
 This research was supported by the Research Executive Agency (REA) of the European Union with 
the  Marie Sk\l{}odowska Curie Individual Fellowship contract numbers 702610 (Resummation4PS, PFM) 
and 659147 (PrecisionTools4LHC, ER) and the 
 ERC Advanced Grant MCatNNLO (340983), by the ERC Consolidator Grant HICCUP (614577), by the ERC Starting Grant PDF4BSM (335260), by the 
UK Science and Technology Facilities Council, and by the Swiss National Science Foundation (SNF) under contracts 200020-175595, 200021-172478
and CRSII2-160814.

\newpage
\appendix

\section{Formulae for the matching schemes}
\label{app:matching}
In this appendix we report the necessary formulae to implement the
matching schemes defined in
Eqs.~\eqref{eq:additive} and~\eqref{eq:multiplicative1} and used in our
study. We start by introducing a convenient notation for the
perturbative expansion of the various ingredients. We define
\begin{align}
\sigma_{\rm tot}^{\rm N^3LO}  = \sum_{i=0}^3\sigma^{(i)},\qquad
\Sigma^{\rm N^3LO}(v) = \sigma^{(0)} + \sum_{i=1}^3\Sigma^{(i)}(v),
\end{align}
where 
\begin{align}
\Sigma^{(i)}(v) = \sigma^{(i)} + \bar{\Sigma}^{(i)}(v),\qquad \bar{\Sigma}^{(i)}(v) \equiv - \int_{v}^{\infty} \rd v' \;\frac{\rd \Sigma^{(i)}(v')}{\rd v'}.
\end{align}
Moreover, we denote the perturbative expansion of the resummed cross
section $\Sigma^{\rm
  N^kLL}$ as
\begin{equation}
\Sigma^{\rm EXP}(v) = \sigma^{(0)} + \sum_{i=1}^3 \Sigma_{\rm N^kLL}^{(i)}(v).
\end{equation}
With this notation, the additive scheme of Eq.~\eqref{eq:additive}
becomes (for simplicity we drop the explicit dependence on $v$ in the following)
\begin{align}
          \Sigma_{\rm add}^{\rm MAT} =& \Sigma^{\rm N^kLL}+ \left\{\sigma^{(1)}+\bar \Sigma^{(1)}- \Sigma_{\rm N^kLL}^{(1)}\right\}+  \left\{\sigma^{(2)}+\bar  \Sigma^{(2)}
                - \Sigma_{\rm N^kLL}^{(2)}\right\} +  \left\{\sigma^{(3)}+\bar  \Sigma^{(3)}
                - \Sigma_{\rm N^kLL}^{(3)}\right\},
\end{align}
where the three terms in curly brackets denote the NLO, NNLO and
N$^3$LO contributions to the matching, respectively.

For the multiplicative scheme we need to introduce the asymptotic
expansion $\Sigma^{\rm N^kLL}_{\rm asym.} $, defined in
Eq.~\eqref{eq:asypt} (the definition for $k\neq 3$ is analogous with
obvious replacements) in terms of the $\tilde{L}\to 0$ limit of the
coefficients $\tilde{\cal L}_{\rm N^kLL}$ of
Eqs.~\eqref{eq:luminosity-NLL},~\eqref{eq:luminosity-NNLL},~\eqref{eq:mod-luminosity-N3LL},
which read
\begin{align}
\label{eq:lumi_asympt}
\tilde{\cal L}_{\rm NLL}^{\tilde{L}\to 0} &= \sum_{c,
  c'}\frac{\rd|\mathcal{M}_{B}|_{cc'}^2}{\rd\Phi_B}
  f_c\!\left(\mu_F,x_1\right)f_{c'}\!\left(\mu_F,x_2\right),\notag\\
\tilde{\cal L}_{\rm NNLL}^{\tilde{L}\to 0} &= \sum_{c,
                                             c'}\frac{\rd|\mathcal{M}_{B}|_{cc'}^2}{\rd\Phi_B}
                                             \sum_{i,
                                             j}\int_{x_1}^{1}\frac{\rd
                                             z_1}{z_1}\int_{x_2}^{1}\frac{\rd
                                             z_2}{z_2}f_i\!\left(\mu_F,\frac{x_1}{z_1}\right)f_{j}\!\left(\mu_F,\frac{x_2}{z_2}\right)\notag\\
&\times\Bigg\{\delta_{ci}\delta_{c'j}\delta(1-z_1)\delta(1-z_2)
\left(1+\frac{\as(\mu_R)}{2\pi} \tilde{H}^{(1)}(\mu_R,x_Q)\right) \notag\\
&+ \frac{\as(\mu_R)}{2\pi}\left(\tilde{C}_{c i}^{(1)}(z_1,\mu_F,x_Q)\delta(1-z_2)\delta_{c'j}+
  \{z_1\leftrightarrow z_2; c,i \leftrightarrow c'j\}\right)\Bigg\},\notag\\
\tilde{\cal L}_{\rm N^3LL}^{\tilde{L}\to 0} & =\sum_{c,
  c'}\frac{\rd|\mathcal{M}_{B}|_{cc'}^2}{\rd\Phi_B} \sum_{i, j}\int_{x_1}^{1}\frac{\rd
  z_1}{z_1}\int_{x_2}^{1}\frac{\rd z_2}{z_2}f_i\!\left(\mu_F,\frac{x_1}{z_1}\right)f_{j}\!\left(\mu_F,\frac{x_2}{z_2}\right)\notag\\&\times\Bigg\{\delta_{ci}\delta_{c'j}\delta(1-z_1)\delta(1-z_2)
\left(1+\frac{\as(\mu_R)}{2\pi} \tilde{H}^{(1)}(\mu_R,x_Q) + \frac{\as^2(\mu_R)}{(2\pi)^2} \tilde{H}^{(2)}(\mu_R,x_Q)\right) \notag\\
&+ \frac{\as(\mu_R)}{2\pi}\left(\tilde{C}_{c i}^{(1)}(z_1,\mu_F,x_Q)\delta(1-z_2)\delta_{c'j}+ \{z_1\leftrightarrow z_2; c,i \leftrightarrow c',j\}\right)\notag\\
& +
  \frac{\as^2(\mu_R)}{(2\pi)^2}\left(\tilde{C}_{c i}^{(2)}(z_1,\mu_F,x_Q)\delta(1-z_2)\delta_{c'j} + \{z_1\leftrightarrow z_2; c,i \leftrightarrow c',j\}\right) \notag\\&+  \frac{\as^2(\mu_R)}{(2\pi)^2}\Big(\tilde{C}_{c i}^{(1)}(z_1,\mu_F,x_Q)\tilde{C}_{c' j}^{(1)}(z_2,\mu_F,x_Q) + G_{c i}^{(1)}(z_1)G_{c' j}^{(1)}(z_2)\Big) \notag\\
& + \frac{\as^2(\mu_R)}{(2\pi)^2} \tilde{H}^{(1)}(\mu_R,x_Q)\Big(\tilde{C}_{c i}^{(1)}(z_1,\mu_F,x_Q)\delta(1-z_2)\delta_{c'j} + \{z_1\leftrightarrow z_2; c,i \leftrightarrow c',j\}\Big) \Bigg\}.
\end{align}
In the following formula the perturbative expansion of $\Sigma^{\rm
  N^kLL}_{\rm asym.}$ is denoted as follows
\begin{equation}
\Sigma^{\rm
  N^kLL}_{\rm asym.} = \sigma^{(0)} + \sum_{i=1}^{k-1}\Sigma_{\rm asym.}^{(i)}.
\end{equation}

With this notation the matching formula~\eqref{eq:multiplicative1} reads
\begin{align}
&	\Sigma_{\rm mult}^{\rm MAT}(v) =  \frac{\Sigma^{\rm N^kLL}}{\Sigma^{\rm
  N^kLL}_{\rm asym.}}\Bigg[\sigma^{(0)}+ \left\{\sigma^{(1)}+\bar \Sigma^{(1)} +\Sigma_{\rm asym.}^{(1)}- \Sigma_{\rm N^kLL}^{(1)} \right\}\nonumber \\ \nonumber \\
	&+ \left\{\sigma^{(2)} + \bar \Sigma^{(2)} +\Sigma_{\rm asym.}^{(2)}- \Sigma_{\rm N^kLL}^{(2)} +\frac{\Sigma_{\rm asym.}^{(1)}}{\sigma^{(0)}}\left(\sigma^{(1)}+\bar \Sigma^{(1)}\right) 
       + \frac{(\Sigma_{\rm N^kLL}^{(1)})^2}{\sigma^{(0)}} - \frac{\Sigma_{\rm N^kLL}^{(1)}}{\sigma^{(0)}}\left(\sigma^{(1)}+\bar \Sigma^{(1)}+\Sigma_{\rm asym.}^{(1)}\right) \right\}\nonumber \\ \nonumber \\
       &+\Bigg\{ \sigma^{(3)} + \bar \Sigma^{(3)} - \Sigma_{\rm N^kLL}^{(3)} 
  - \frac{(\Sigma_{\rm N^kLL}^{(1)})^3}{(\sigma^{(0)})^2} + \frac{(\Sigma_{\rm N^kLL}^{(1)})^2}{(\sigma^{(0)})^2}\left(\sigma^{(1)}+\bar \Sigma^{(1)} + \Sigma_{\rm asym.}^{(1)}\right) \nonumber \\
  &+\frac{1}{\sigma_0}\left((\sigma^{(1)}+\bar \Sigma^{(1)})(\Sigma_{\rm asym.}^{(2)}-\Sigma_{\rm N^kLL}^{(2)}) 
 +\Sigma_{\rm asym.}^{(1)}(\sigma^{(2)}+\bar \Sigma^{(2)} - \Sigma_{\rm N^kLL}^{(2)})\right) \nonumber\\
 &- \frac{1}{(\sigma^{(0)})^2}\Sigma_{\rm N^kLL}^{(1)}\left(\Sigma_{\rm asym.}^{(1)}(\sigma^{(1)}+\bar \Sigma^{(1)}) + \sigma^{(0)}(\sigma^{(2)}+\bar \Sigma^{(2)} + \Sigma_{\rm asym.}^{(2)}- 2\Sigma_{\rm N^kLL}^{(2)})\right) \Bigg\}\Bigg],
\end{align}
where, as above, we grouped the terms entering at NLO, NNLO, and
N$^3$LO within curly brackets.

%===================================================================================================
\section{Formulae for \texorpdfstring{N$^3$LL}{N3LL} resummation}
\label{app:sudakov-radiator}

In this section we report the expressions for quantities needed for
N$^3$LL resummation of transverse observables, that we have used
throughout this article.

\vspace{3mm}
First of all we report our convention for the RG equation of the
strong coupling which reads
\begin{equation}
  \frac{\rd\as(\mu)}{\rd\ln \mu^2}
  =
  \beta(\as) \equiv
  -\as\left( \beta_0 \as +\beta_1\as^2 +\beta_2
  \as^3 +\beta_3 \as^4 + \dots\right),
\end{equation}
where the coefficients of the $\beta$-function are
\begin{eqnarray}
  \beta_0 &=& \frac{11 C_A - 2 n_f}{12\pi}\,,\qquad 
  \beta_1 = \frac{17 C_A^2 - 5 C_A n_f - 3 C_F n_f}{24\pi^2}\,,\\
  \beta_2 &=& \frac{2857 C_A^3+ (54 C_F^2 -615C_F C_A -1415 C_A^2)n_f
       +(66 C_F +79 C_A) n_f^2}{3456\pi^3}\,,\\
\beta_3 &=& \frac{1}{(4\pi)^4}\Bigg\{C_A C_F n_f^2 \frac14\left(\frac{17152}{243} + \frac{448}9 \zeta_3\right) + 
C_A C_F^2 n_f \frac12\left(-\frac{4204}{27} + \frac{352}{9} \zeta_3\right)\nonumber\\
&&\hspace{10mm} + \frac{53}{243} C_A n_f^3 + C_A^2 C_F n_f\frac12 \left(\frac{7073}{243} - \frac{656}9 \zeta_3\right) + 
C_A^2 n_f^2 \frac14\left(\frac{7930}{81} + \frac{224}9 \zeta_3\right)\nonumber\\
&&\hspace{10mm} + \frac{154}{243} C_F n_f^3 + 
C_A^3 n_f \frac12\left(-\frac{39143}{81} + \frac{136}3 \zeta_3\right) + C_A^4 \left(\frac{150653}{486} - \frac{44}9 \zeta_3\right)\nonumber\\
&&\hspace{10mm} + C_F^2 n_f^2 \frac14\left(\frac{1352}{27} - \frac{704}9 \zeta_3 \right) + 23 C_F^3 n_f + n_f \frac{d_F^{abcd}d_A^{abcd}}{N_A} \left(\frac{512}9 - \frac{1664}3 \zeta_3\right)\nonumber\\
&&\hspace{10mm} + n_f^2\frac{d_F^{abcd}d_F^{abcd}}{N_A} \left(-\frac{704}9 + \frac{512}3 \zeta_3\right) + \frac{d_A^{abcd}d_A^{abcd}}{N_A} \left(-\frac{80}9 + \frac{704}3 \zeta_3\right)\Bigg\}\,,
\end{eqnarray}
with
\begin{eqnarray*}
\frac{d_F^{abcd}d_F^{abcd}}{N_A} = \frac{N_c^4 - 6 N_c^2 + 18}{96 N_c^2},\qquad
\frac{d_F^{abcd}d_A^{abcd}}{N_A} = \frac{N_c(N_c^2 + 6)}{48},\qquad
\frac{d_A^{abcd}d_A^{abcd}}{N_A} = \frac{N_c^2 (N_c^2 + 36)}{24},
\end{eqnarray*}
and $C_A = N_c$, $C_F = \frac{N_c^2-1}{2N_c}$, and $N_c = 3$.

\vspace{3mm} We also provide expressions for the functions
$g_i(\lambda)$ entering in the N$^3$LL Sudakov radiator
Eq.~\eqref{eq:mod-radiator} and its derivative. We define
\begin{equation}
  \lambda = \as(\mu_R) \beta_0 \tilde L \,.
\end{equation}
We have:
\begin{align}
  g_{1}(\lambda) =& \frac{A^{(1)}}{\pi\beta_{0}}\frac{2 \lambda +\ln (1-2 \lambda )}{2  \lambda }, \\
  g_{2}(\lambda) =& \frac{1}{2\pi \beta_{0}}\ln (1-2 \lambda )
  \left(A^{(1)} \ln \frac{1}{x_Q^2}+B^{(1)}\right)
  -\frac{A^{(2)}}{4 \pi ^2 \beta_{0}^2}\frac{2 \lambda +(1-2
    \lambda ) \ln (1-2 \lambda )}{1-2
    \lambda} \notag\\
  &+A^{(1)} \bigg(-\frac{\beta_{1}}{4 \pi \beta_{0}^3}\frac{\ln
    (1-2 \lambda ) ((2 \lambda -1) \ln (1-2 \lambda )-2)-4
    \lambda}{1-2 \lambda}\notag\\
  &\hspace{10mm}-\frac{1}{2 \pi \beta_{0}}\frac{(2 \lambda(1
    -\ln (1-2 \lambda ))+\ln (1-2 \lambda ))}{1-2\lambda} \ln
  \frac{\mu_R^2}{x_Q^2 M^2}\bigg)\,,\\
  g_{3}(\lambda) =
  & \left(A^{(1)} \ln\frac{1}{x_Q^2}+B^{(1)}\right)
  \bigg(-\frac{\lambda }{1-2 \lambda} \ln
  \frac{\mu _{R}^2}{x_Q^2M^2}+\frac{\beta_{1}}{2 \beta_{0}^2}\frac{2 \lambda
    +\ln (1-2 \lambda )}{1-2 \lambda}\bigg)\notag\\
  &   -\frac{1}{2 \pi\beta_{0}}\frac{\lambda}{1-2\lambda}\left(A^{(2)}
  \ln\frac{1}{x_Q^2}+B^{(2)}\right)-\frac{A^{(3)}}{4 \pi ^2 \beta_{0}^2}\frac{\lambda ^2}{(1-2\lambda )^2} \notag\\
  &   +A^{(2)} \bigg(\frac{\beta_{1}}{4 \pi  \beta_{0}^3 }\frac{2 \lambda  (3
    \lambda -1)+(4 \lambda -1) \ln (1-2 \lambda )}{(1-2 \lambda
    )^2}-\frac{1}{\pi \beta_{0}}\frac{\lambda ^2 }{(1-2 \lambda )^2}\ln\frac{\mu_R^2}{x_Q^2 M^2}\bigg) \notag\\
  & +A^{(1)} \bigg(\frac{\lambda  \left(\beta_{0} \beta_{2} (1-3 \lambda
    )+\beta_{1}^2 \lambda \right)}{\beta_{0}^4 (1-2 \lambda)^2}
  +\frac{(1-2 \lambda) \ln (1-2 \lambda ) \left(\beta_{0} \beta_{2} 
    (1-2 \lambda )+2 \beta_{1}^2 \lambda \right)}{2\beta_{0}^4 (1-2 \lambda)^2} 
  \notag\\
  &\hspace{10mm}+\frac{\beta_{1}^2}{4 \beta_{0}^4}
  \frac{(1-4 \lambda ) \ln ^2(1-2 \lambda )}{(1-2 \lambda)^2}-\frac{\lambda ^2 }{(1-2 \lambda
    )^2} \ln ^2\frac{\mu_R^2}{x_Q^2 M^2}\notag\\
  &
  \hspace{10mm}   -\frac{\beta_{1}}{2 \beta_{0}^{2}}\frac{(2 \lambda  (1-2 \lambda)+(1-4 \lambda) \ln (1-2 \lambda ))
  }{(1-2\lambda )^2}\ln\frac{\mu_R^2}{x_Q^2 M^2}\bigg)\,,\\
  g_4(\lambda)  =& \frac{A^{(4)} (3-2 \lambda ) \lambda ^2}{24 \pi ^2 \beta_0^2 (2 \lambda -1)^3}\notag\\
  & + \frac{A^{(3)}}{48 \pi 
    \beta_0^3 (2 \lambda -1)^3}\Bigg\{3 \beta_1 (1-6 \lambda ) \ln (1-2 \lambda )+2 \lambda  \Bigg(\beta_1 (5 \lambda  (2 \lambda -3)+3)\notag\\
  &\hspace{10mm} +6 \beta_0^2 (3-2 \lambda ) \lambda  \ln
  \frac{\mu_R^2}{x_Q^2M^2}\Bigg)+12 \beta_0^2 (\lambda -1) \lambda
  (2 \lambda -1) \ln \frac{1}{x_Q^2}\Bigg\} \notag\\
  & + \frac{A^{(2)}}{24
    \beta_0^4 (2 \lambda -1)^3} \Bigg\{32 \beta_0 \beta_2 \lambda ^3-2 \beta_1^2 \lambda 
  (\lambda  (22 \lambda -9)+3)\notag\\
  &\hspace{10mm}+12 \beta_0^4 (3-2 \lambda ) \lambda ^2 \ln
  ^2\frac{\mu_R^2}{x_Q^2M^2}+6 \beta_0^2 \ln
  \frac{\mu_R^2}{x_Q^2M^2}\times\notag\\
  &\hspace{10mm}\left(\beta_1 (1-6 \lambda ) \ln (1-2
  \lambda )+2 (\lambda -1) \lambda  (2 \lambda -1) \left(\beta_1+2 \beta_0^2 \ln\frac{1}{x_Q^2}\right)\right)\notag\\
  &\hspace{10mm}+3 \beta_1 \Bigg(\beta_1 \ln (1-2
  \lambda ) (2 \lambda +(6 \lambda -1) \ln (1-2 \lambda )-1)\notag\\
  &\hspace{10mm}-2 \beta_0^2 (2 \lambda -1) (2
  (\lambda -1) \lambda -\ln (1-2 \lambda )) \ln \frac{1}{x_Q^2}\Bigg)\Bigg\}\notag\notag\\
  & + \frac{\pi  A^{(1)}}{12 \beta_0^5 (2 \lambda -1)^3} \Bigg\{\beta_1^3 (1-6 \lambda ) \ln ^3(1-2 \lambda )+3 \ln (1-2 \lambda )
  \Bigg(\beta_0^2 \beta_3 (2 \lambda -1)^3\notag\\
  &\hspace{10mm}+\beta_0 \beta_1
  \beta_2 \left(1-2 \lambda  \left(8 \lambda ^2-4 \lambda +3\right)\right)+4 \beta_1^3
  \lambda ^2 (2 \lambda +1)\notag\\
  &\hspace{10mm}+\beta_0^2 \beta_1 \ln \frac{\mu_R^2}{
    x_Q^2M^2} \left(\beta_0^2 (1-6 \lambda ) \ln \frac{\mu_R^2}{
    x_Q^2M^2}-4 \beta_1 \lambda \right)\Bigg)\notag\\
  &\hspace{10mm}+3 \beta_1^2 \ln ^2(1-2 \lambda
  ) \left(2 \beta_1 \lambda +\beta_0^2 (6 \lambda -1) \ln \frac{\mu_R^2}{x_Q^2M^2}\right)\notag\\
  &\hspace{10mm}+3 \beta_0^2 (2 \lambda -1) \ln
  \frac{1}{x_Q^2} \Bigg(-\beta_1^2 \ln ^2(1-2 \lambda ) +2 \beta_0^2
  \beta_1 \ln (1-2 \lambda ) \ln \frac{\mu_R^2}{
    x_Q^2M^2}\notag\\
  &\hspace{10mm}+4 \lambda 
  \left(\lambda  \left(\beta_1^2-\beta_0 \beta_2\right)+\beta_0^4
  (\lambda -1) \ln ^2\frac{\mu_R^2}{x_Q^2M^2}\right)\Bigg)\notag\\
  &\hspace{10mm}+2 \lambda  \Bigg(\beta_0^2 \beta_3 ((15-14 \lambda )
  \lambda -3)+\beta_0 \beta_1 \beta_2 (5 \lambda  (2 \lambda -3)+3)\notag\\
  &\hspace{10mm}+4
  \beta_1^3 \lambda ^2+2 \beta_0^6 (3-2 \lambda ) \lambda  \ln
  ^3\frac{\mu_R^2}{x_Q^2M^2}+3 \beta_0^4 \beta_1 \ln
  ^2\frac{\mu_R^2}{x_Q^2M^2}\notag\\
  &\hspace{10mm}+6 \beta_0^2 \lambda  (2 \lambda +1)
  \left(\beta_0 \beta_2-\beta_1^2\right) \ln \frac{\mu_R^2}{x_Q^2M^2}-8 \beta_0^6 \left(4 \lambda ^2-6 \lambda +3\right) \zeta_3\Bigg)\Bigg\}\notag\\
  & + \frac{B^{(3)} (\lambda -1) \lambda }{4 \pi  \beta_0 (1-2 \lambda )^2}+ \frac{B^{(2)} \left(\beta_1 \ln (1-2 \lambda )-2 (\lambda -1) \lambda  \left(\beta_1-2 \beta_0^2 \ln \frac{\mu_R^2}{x_Q^2M^2}\right)\right)}{4\beta_0^2
    (1-2\lambda )^2}\notag\\
  & + \frac{\pi  B^{(1)}}{4 \beta_0^3 (1-2 \lambda )^2} \Bigg\{4 \lambda  \left(\lambda 
  \left(\beta_1^2-\beta_0 \beta_2\right)+\beta_0^4 (\lambda -1) \ln
  ^2\frac{\mu_R^2}{x_Q^2M^2}\right)\notag\\
  &\hspace{10mm}-\beta_1^2 \ln ^2(1-2 \lambda )+2 \beta_0^2 \beta_1
  \ln (1-2 \lambda ) \ln \frac{\mu_R^2}{x_Q^2M^2}\Bigg\}.
\end{align}
For Higgs boson production in gluon fusion, the coefficients $A^{(i)}$
and $B^{(i)}$ which enter the formulae above
are (in units of $\as/(2\pi)$)
\begin{align}
  A_{\rm ggH}^{(1)} =& \,2 C_A,
  \notag\\
  \vspace{1.5mm}
  A_{\rm ggH}^{(2)} =&
  \left( \frac{67}{9}-\frac{\pi ^2}{3} \right) C_A^2
  -\frac{10}{9} C_A n_f,
  \notag\\
  \vspace{1.5mm}
  A_{\rm ggH}^{(3)} =&
   \left( -22 \zeta_3 - \frac{67 \pi^2}{27}+\frac{11 \pi^4}{90}+\frac{15503}{324} \right) C_A^3
  + \left( \frac{10 \pi^2}{27}-\frac{2051}{162} \right) C_A^2 n_f\notag\\
  &+ \left( 4 \zeta_3-\frac{55}{12} \right) C_A C_F n_f
  + \frac{50}{81} C_A n_f^2,
  \notag\\
  \vspace{1.5mm}
  A_{\rm ggH}^{(4)} =&
     \left( \frac{121}{3} \zeta_3 \zeta_2-\frac{8789 \zeta_2}{162}-\frac{19093 \zeta_3}{54}-\frac{847 \zeta_4}{24}+132 \zeta_5+\frac{3761815}{11664} \right) C_A^4
   + \left( -\frac{4 \zeta_3}{9}-\frac{232}{729} \right) C_A n_f^3
  \notag\\&
   + \left( -\frac{22}{3} \zeta_3 \zeta_2+\frac{2731 \zeta_2}{162}+\frac{4955 \zeta_3}{54}+\frac{11 \zeta_4}{6}-24 \zeta_5-\frac{31186}{243} \right) C_A^3 n_f\notag\\
&   + \left( -\frac{38 \zeta_3}{9}-2 \zeta_4+\frac{215}{24} \right) C_A C_F n_f^2
   + \left( \frac{272 \zeta_3}{9}+11 \zeta_4-\frac{7351}{144} \right) C_A^2 C_F n_f\notag\\
&   + \left( -\frac{103 \zeta_2}{81}-\frac{47 \zeta_3}{27}+\frac{5
  \zeta_4}{6}+\frac{13819}{972} \right) C_A^2 n_f^2 +
  \Gamma^{(4)}_{\rm cusp, ggH} + C_A \Delta {\rm A}^{(4)},
  \notag\\
  \vspace{1.5mm}
  B_{\rm ggH}^{(1)} =&
  -\frac{11}{3} C_A + \frac{2}{3}n_f,
  \notag\\
  \vspace{1.5mm}
  B_{\rm ggH}^{(2)} =&
  \left( \frac{11 \zeta _2}{6}-6 \zeta _3-\frac{16}{3} \right) C_A^2 
  + \left( \frac{4}{3}-\frac{\zeta _2}{3} \right) C_A n_f 
  + n_f C_F,
  \notag\\
  \vspace{1.5mm}
  B_{\rm ggH}^{(3)} =&
  \left( \frac{22 \zeta _3 \zeta _2}{3}-\frac{799 \zeta _2}{81}-\frac{5 \pi ^2 \zeta _3}{9}-\frac{2533 \zeta _3}{54}-\frac{77 \zeta _4}{12}+20 \zeta _5-\frac{319 \pi ^4}{1080}+\frac{6109 \pi
   ^2}{1944}+\frac{34219}{1944} \right) C_A^3
  \notag\\&
  + \left( \frac{103 \zeta _2}{81}+\frac{202 \zeta _3}{27}-\frac{5 \zeta _4}{6}+\frac{41 \pi ^4}{540}-\frac{599 \pi ^2}{972}-\frac{10637}{1944} \right) C_A^2 n_f\notag\\
&  + \left( -\frac{2 \zeta _3}{27}+\frac{5 \pi ^2}{162}+\frac{529}{1944} \right) C_A n_f^2
   + \left( 2 \zeta _4-\frac{\pi ^4}{45}-\frac{\pi ^2}{12}+\frac{241}{72} \right) C_A C_F n_f\notag\\
&  - \frac{1}{4} C_F^2 n_f
  - \frac{11}{36} C_A n_f^2+ C_A \Delta {\rm B}^{(3)}.
\end{align}

For Drell-Yan production, the coefficients read
\begin{align}
  A_{\rm DY}^{(1)} =& 2C_F,
  \notag\\
  \vspace{1.5mm}
  A_{\rm DY}^{(2)} =& \left( \frac{67}{9} - \frac{\pi^2}{3} \right)C_A C_F - \frac{10}{9}C_F n_f,
  \notag\\
  \vspace{1.5mm}
  A_{\rm DY}^{(3)} =&
  \left( \frac{15503}{324} - \frac{67\pi^2}{27} + \frac{11\pi^4}{90} - 22\zeta_3 \right)C_A^2 C_F
  + \left( -\frac{2051}{162} + \frac{10\pi^2}{27} \right)C_A C_F n_f
  \notag\\&
  + \left( -\frac{55}{12} + 4\zeta_3 \right) C_F^2 n_f
  + \frac{50}{81} C_F n_f^2,
  \notag\\
  \vspace{1.5mm}
  A_{\rm DY}^{(4)} =&
    \left( \frac{3761815}{11664} - \frac{8789\zeta_2}{162} -
                      \frac{19093\zeta_3}{54} +
                      \frac{121\zeta_2\zeta_3}{3} -
                      \frac{847\zeta_4}{24} + 132\zeta_5 \right) C_A^3
                      C_F\notag\\
&  + \left(-\frac{232}{729} -\frac{4\zeta_3}{9} \right) C_F n_f^3 + \left( \frac{215}{24} - \frac{38\zeta_3}{9} - 2\zeta_4 \right) C_F^2 n_f^2\notag\\
&  + \left(-\frac{31186}{243} + \frac{2731\zeta_2}{162} + \frac{4955\zeta_3}{54} - \frac{22\zeta_2\zeta_3}{3} + \frac{11\zeta_4}{6} - 24\zeta_5 \right) C_A^2 C_F n_f
  \notag\\&
  + \left(-\frac{7351}{144} + \frac{272\zeta_3}{9} + 11\zeta_4  \right) C_A C_F^2 n_f
  + \left( \frac{13819}{972} - \frac{103\zeta_2}{81} -
            \frac{47\zeta_3}{27} + \frac{5\zeta_4}{4} \right) C_A C_F
            n_f^2\notag\\
&+\Gamma^{(4)}_{\rm cusp, DY}+ C_F \Delta {\rm A}^{(4)},
  \notag\\
  \vspace{1.5mm}
  B_{\rm DY}^{(1)} =&
  -3 C_F,
  \notag\\
  \vspace{1.5mm}
  B_{\rm DY}^{(2)} =&
  \left(-\frac{17}{12} - \frac{11 \pi^2}{12} +  6 \zeta_3\right) C_A C_F+
  \left( -\frac{3}{4} + \pi^2 - 12\zeta_3 \right) C_F^2+
  \left( \frac{1}{6}+\frac{\pi^2}{6} \right) C_F n_f,
  \notag\\
  \vspace{1.5mm}
  B_{\rm DY}^{(3)} =&
    \left( \frac{22 \zeta _3 \zeta _2}{3}-\frac{799 \zeta _2}{81}-\frac{11 \pi ^2 \zeta _3}{9}+\frac{2207 \zeta _3}{54}-\frac{77 \zeta _4}{12}-10 \zeta
   _5-\frac{83 \pi ^4}{360}-\frac{7163 \pi ^2}{1944}+\frac{151571}{3888} \right) C_A^2 C_F
  \notag\\&
  + \left( \frac{4 \pi ^2 - 51}{3}\zeta_3 + 60 \zeta _5-\frac{2 \pi ^4}{5}-\frac{3 \pi ^2}{4}-\frac{29}{8} \right) C_F^3
  + \left( \frac{34 \zeta _3}{3}+2 \zeta _4-\frac{7 \pi ^4}{54}-\frac{13 \pi ^2}{36}+\frac{23}{4} \right) C_F^2 n_f
  \notag\\
&  + \left( -\frac{2}{3} \pi ^2 \zeta _3-\frac{211 \zeta _3}{3}-30
            \zeta _5+\frac{247 \pi ^4}{540}+\frac{205 \pi
            ^2}{36}-\frac{151}{16} \right) C_A C_F^2\notag\\
&  + \left( \frac{103 \zeta _2}{81}-\frac{128 \zeta _3}{27}-\frac{5 \zeta _4}{6}+\frac{11 \pi ^4}{180}+\frac{1297 \pi ^2}{972}-\frac{3331}{243} \right) C_A C_F n_f\notag\\
 & + \left( \frac{10 \zeta _3}{27}-\frac{5 \pi ^2}{54}+\frac{1115}{972} \right) C_F n_f^2
+ C_F \Delta {\rm B}^{(3)}.
\end{align}
The expressions for the coefficients $A^{(i)}$ and $B^{(i)}$ are
extracted from
Refs.~\cite{deFlorian:2001zd,Becher:2012yn,Li:2016ctv,Vladimirov:2016dll}
for Higgs boson production and
Refs.~\cite{Davies:1984hs,Becher:2010tm,Li:2016ctv,Vladimirov:2016dll}
for DY production. The N$^3$LL anomalous dimension $A^{(4)}$ receives
a contribution from the four-loop cusp anomalous dimension
$\Gamma^{(4)}_{\rm cusp}$, that has recently been computed numerically
in ref.~\cite{Moch:2018wjh}, and is given by
\begin{align}
\label{eq:gamma4}
\Gamma^{(4)}_{\rm cusp, ggH} &\simeq 2555 -732.125\, n_f +27.5031\, n_f^2 + 0.460173\, n_f^3\,,\notag\\
\Gamma^{(4)}_{\rm cusp, DY} &\simeq  1293.88 -323.244\, n_f + 12.2236\, n_f^2+0.204522\, n_f^3\,.
\end{align}
The extra terms
\begin{equation}
\Delta{\rm A}^{(4)} = -64 \pi^3 \beta_0^3 \zeta_3 ,\qquad \Delta
{\rm B}^{(3)} = -32 \pi^2 \beta_0^2 \zeta_3, \qquad \Delta {\rm H}^{(2)} =\frac{16}{3} \pi
  \beta_0 \zeta_3,
\end{equation}
are a feature of performing the resummation in momentum space, and do
not appear in the anomalous dimensions in $b$ space (see
Ref.~\cite{Bizon:2017rah} for details). The term
$\Delta {\rm H}^{(2)} $ will appear in the $\tilde{H}$ functions
defined below.

We also present the expansion of hard-virtual coefficient function $H$
in powers of the strong coupling
\begin{equation}
  H(M) =
  1
  +
  \sum_{n=1}^{2} \left( \frac{\as(M)}{2\pi} \right)^n \, H^{(n)}(M),
\end{equation}
with
\begin{align}
\label{eq:H-fun-G}
  H_{{\rm ggH}}^{(1)}(M) =&  C_A\left(5+\frac{7}{6}\pi^2\right)-3 C_F,\notag\\
  H_{{\rm ggH}}^{(2)}(M) =&   \frac{5359}{54} + \frac{137}{6}\ln\frac{m_H^2}{m_t^2} 
+ \frac{1679}{24}\pi^2 + \frac{37}{8}\pi^4- \frac{499}{6}\zeta_3
              + C_A \Delta {\rm H}^{(2)} \,,\qquad n_f=5,
\end{align}
and
\begin{align}
\label{eq:H-fun-Q}
  H_{{\rm DY}}^{(1)}(M) =& C_F \left( 5 + \frac{7}{6}\pi^2 \right), \notag\\
  H_{{\rm DY}}^{(2)}(M) =&  -\frac{57433}{972}+\frac{281}{162}\pi^2
               +\frac{22}{27}\pi^4 +\frac{1178}{27}\zeta_3+ C_F \Delta {\rm H}^{(2)} \,,\qquad n_f=5.
\end{align}
Their renormalisation-scale dependence is given by
\begin{align}
H^{(1)}(\mu_R) &= H^{(1)}(M) + 2 d_B \pi \beta_0 \ln\frac{\mu_R^2}{M^2},\\
H^{(2)}(\mu_R) &= H^{(2)}(M) + 4  d_B \left( \frac{1+d_{B} }{2} \pi^2\beta_0^2 \ln^2\frac{\mu_R^2}{M^2} + \pi^2 \beta_1
  \ln\frac{\mu_R^2}{M^2}\right)\notag\\
& + 2 \left(1+d_{B}\right) \pi\beta_0  \ln\frac{\mu_R^2}{M^2} H^{(1)}(M),
\end{align}
where $d_B$ is the strong-coupling order of the Born squared
amplitude  (e.g. $d_B=2$ for Higgs production).
The factors $\tilde{H}$ that appear in the luminosity prefactors
(Eqs.~\eqref{eq:luminosity-NLL},~\eqref{eq:luminosity-NNLL},~\eqref{eq:mod-luminosity-N3LL})
are defined as
\begin{align}
\tilde{H}^{(1)}(\mu_R,&x_Q) = H^{(1)}(\mu_R) +
                             \left(-\frac{1}{2}A^{(1)}\ln x_Q^2 +
                             B^{(1)}\right) \ln x_Q^2,\notag\\
\tilde{H}^{(2)}(\mu_R,&x_Q) = H^{(2)}(\mu_R) +
                             \frac{(A^{(1)})^2}{8}\ln^4x_Q^2 - \left(\frac{A^{(1)}
                             B^{(1)}}{2}+\frac{A^{(1)}}{3}\pi\beta_0
                             \right)\ln^3 x_Q^2\notag\\
&+\left(\frac{-A^{(2)}+(B^{(1)})^2}{2} + \pi\beta_0
  \left(B^{(1)}+A^{(1)}\ln \frac{x_Q^2 M^2}{\mu_R^2}\right)\right)\ln^2 x_Q^2\notag\\
& - \left(-B^{(2)}+B^{(1)}2\pi\beta_0\ln \frac{x_Q^2
  M^2}{\mu_R^2}\right)\ln x_Q^2  + H^{(1)}(\mu_R)\ln x_Q^2\left( -\frac{1}{2}A^{(1)}\ln x_Q^2 +
                             B^{(1)} \right).
\end{align}
Finally we report the expansion of the collinear coefficient functions $C_{ab}$
\begin{align}
  C_{ab}(z) =& \delta(1-z)\delta_{ab} + \sum_{n=1}^{2} \left( \frac{\as(\mu)}{2\pi} \right)^n \,C_{ab}^{(n)}(z),
\end{align}
where $\mu$ is the same scale that enters parton densities. The
first-order expansion has been known for a long time and reads 
\begin{equation}
\label{eq:coeff-fun}
C_{ab}^{(1)}(z)= - \hat P_{ab}^{(0),\epsilon}(z) - \delta_{ab}\delta(1-z)\frac{\pi^2}{12}\,C ,
\end{equation}
with $C=C_A,\,C_F$ for the $gg$ and $qq$ case, respectively.
$\hat P_{ab}^{(0),\epsilon}(z)$ is the $\mathcal{O}(\epsilon)$ part of
the leading-order regularised splitting functions
$\hat P_{ab}^{(0)}(z)$
\begin{align}
  &\hat P^{(0)}_{qq}(z)=C_F\left[\frac{1+z^2}{(1-z)_+}+\frac32\delta(1-z)\right],  &\hat P^{(0),\epsilon}_{qq}(z) = -C_F (1-z), \nonumber\\
  &\hat P^{(0)}_{qg}(z)=\frac12\left[z^2+(1-z)^2\right],   &\hat P^{(0),\epsilon}_{qg}(z) = -z(1-z),\ \ \ \nonumber\\
  &\hat P^{(0)}_{gq}(z)=C_F\frac{1+(1-z)^2}{z},   &\hat P^{(0),\epsilon}_{gq}(z) = -C_F z,\ \ \ \ \ \ \ \ \nonumber\\
  &\hat P^{(0)}_{gg}(z)=2C_A\left[\frac z{(1-z)_+}+\frac{1-z}z+z(1-z)\right]+2\pi\beta_0\delta(1-z),  & \hat P^{(0),\epsilon}_{gg}(z) = 0.\ \ \ \ \ \ \ \ \ \qquad
\end{align}
The second-order collinear coefficient functions $C_{ab}^{(2)}(z)$, as
well as the $G$ coefficients (see Eqs.~\eqref{eq:luminosity-NLL},
\eqref{eq:luminosity-NNLL}, \eqref{eq:mod-luminosity-N3LL}) for
gluon-fusion processes are obtained in
Refs.~\cite{Catani:2011kr,Gehrmann:2014yya,Echevarria:2016scs}, while
for quark-induced processes they are derived in
Ref.~\cite{Catani:2012qa}. In the present work we extract their
expressions using the results of
Refs.~\cite{Catani:2011kr,Catani:2012qa}. For gluon-fusion processes,
the $C^{(2)}_{gq}$ and $C^{(2)}_{gg}$ coefficients normalised as in
Eq.~\eqref{eq:coeff-fun} are extracted from Eqs.~(30) and~(32) of
Ref.~\cite{Catani:2011kr}, respectively, where we use the hard
coefficients of Eqs.~\eqref{eq:H-fun-G} {\it without} the new term
$\Delta {\rm H}^{(2)}$ in the $H_g^{(2)}(M)$
coefficient.\footnote{These must be replaced by $H^{(1)}\to H^{(1)}/2$
  and $H^{(2)}\to H^{(2)}/4$ to match the convention of
  Refs.~\cite{Catani:2011kr,Catani:2012qa}.} The coefficient $G^{(1)}$
is taken from Eq.~(13) of Ref.~\cite{Catani:2011kr}. Similarly, for
quark-initiated processes, we extract $C^{(2)}_{qg}$ and
$C^{(2)}_{qq}$ from Eqs.~(32) and~(34) of Ref.~\cite{Catani:2012qa},
respectively, where we use the hard coefficients from
Eqs.~\eqref{eq:H-fun-Q} {\it without} the new term
$\Delta {\rm H}^{(2)}$ in the $H_q^{(2)}(M)$ coefficient. The
remaining quark coefficient function $C^{(2)}_{q\bar{q}}$,
$C^{(2)}_{q\bar{q}'}$ and $C^{(2)}_{qq'}$ are extracted from Eq.~(35)
of the same article.

The coefficients $\tilde{C}$ in
Eqs.~\eqref{eq:luminosity-NLL},~\eqref{eq:luminosity-NNLL},~\eqref{eq:mod-luminosity-N3LL}
are defined as
\begin{align}
\tilde{C}_{ab}^{(1)}(z,&\mu_F,x_Q) = C_{ab}^{(1)}(z) +
                                    \hat{P}_{ab}^{(0)}(z)\ln\frac{x_Q^2 M^2}{\mu_F^2},\notag\\
\tilde{C}_{ab}^{(2)}(z,&\mu_F,x_Q) = C_{ab}^{(2)}(z) +
                                    \pi\beta_0 \hat{P}_{ab}^{(0)}(z)\left(
                                    \ln^2\frac{x_Q^2 M^2}{\mu_F^2} -
                                   2 \ln\frac{x_Q^2 M^2}{\mu_F^2}
                                    \ln\frac{x_Q^2
                                    M^2}{\mu_R^2}\right) +
                                    \hat{P}_{ab}^{(1)}(z)\ln\frac{x_Q^2
                                    M^2}{\mu_F^2} \notag\\
& + \frac{1}{2}(\hat{P}^{(0)}\otimes \hat{P}^{(0)})_{ab}(z) \ln^2\frac{x_Q^2
  M^2}{\mu_F^2} + (C^{(1)}\otimes \hat{P}^{(0)})_{ab}(z) \ln\frac{x_Q^2
  M^2}{\mu_F^2} - 2\pi\beta_0 C_{ab}^{(1)}(z) \ln\frac{x_Q^2
  M^2}{\mu_R^2}.
\end{align}

\bibliographystyle{JHEP}
\bibliography{pheno}

\end{document}